\documentclass[twocolumn]{aastex62}
\usepackage{graphicx}
\usepackage{amssymb}
\usepackage{gensymb}
\usepackage[caption=false]{subfig}
\usepackage{booktabs}
\usepackage{xcolor}
\usepackage{multirow}

\newcommand{\teff}{$T_{\rm{eff}}$}

\newcommand{\TOIOneMs}{0.312$\pm$0.007} 
\newcommand{\TOIOneRsMann}{0.334$\pm$0.010} 
\newcommand{\TOIOneTeff}{3403$\pm$100} 
\newcommand{\TOIOneLsun}{0.0140$\pm$0.0003} 
\newcommand{\TOIOneDist}{62.23$\pm$0.21} 
\newcommand{\TOIOneRa}{22:11:47.300} 
\newcommand{\TOIOneDec}{-58:56:42.25} 
\newcommand{\TOIOneRv}{-72.4$\pm$1.0} 
\newcommand{\TOIOneLCDens}{12.8$^{+9.5}_{-4.2}$} 
\newcommand{\TOIOneMannDens}{11.8$\pm$2.0} 
\newcommand{\TOIOnePMRA}{138.138$\pm$0.089} 
\newcommand{\TOIOnePMDEC}{-235.81$\pm$0.076} 

\newcommand{\TOIOnePeriod}{5.078030$\pm$0.000015} 
\newcommand{\TOIOneMp}{8.8$^{+9.0}_{-3.1}$} 
\newcommand{\TOIOneRp}{2.72$\pm$0.18} 
\newcommand{\TOIOneRpRs}{0.075$\pm$0.003} 
\newcommand{\TOIOneaRs}{25.2$\pm$1.5} 
\newcommand{\TOIOneaRsLC}{25.9$^{+5.3}_{-3.2}$} 
\newcommand{\TOIOnei}{88.4$^{+0.6}_{-0.4}$} 
\newcommand{\TOIOneb}{0.72$^{+0.07}_{-0.18}$} 
\newcommand{\TOIOneIns}{8.8$\pm$1.0} 
\newcommand{\TOIOneTeqUpper}{471} 
\newcommand{\TOIOneTeqHigh}{431} 
\newcommand{\TOIOneTeqLow}{333} 

\newcommand{\TOITwoMs}{0.179$\pm$0.004} 
\newcommand{\TOITwoRsMann}{0.211$\pm$0.006} 
\newcommand{\TOITwoTeff}{3212$\pm$100} 
\newcommand{\TOITwoLsun}{0.0041$\pm$0.0001} 
\newcommand{\TOITwoDist}{38.11$\pm$0.23} 
\newcommand{\TOITwoRa}{23:32:58.270} 
\newcommand{\TOITwoDec}{-29:24:54.19} 
\newcommand{\TOITwoRv}{7.8$\pm$1.0} 
\newcommand{\TOITwoLCDens}{25.6$^{+4.3}_{-8.7}$} 
\newcommand{\TOITwoMannDens}{27.0$\pm$4.0} 
\newcommand{\TOITwoPMRA}{151.047$\pm$0.108} 
\newcommand{\TOITwoPMDEC}{-333.194$\pm$0.156} 

\newcommand{\TOITwoPeriod}{5.436098$\pm$0.000039} 
\newcommand{\TOITwoMp}{3.0$^{+2.0}_{-1.1}$} 
\newcommand{\TOITwoRp}{1.44$\pm0.12$} 
\newcommand{\TOITwoRpRs}{0.062$\pm$0.002} 
\newcommand{\TOITwoaRs}{34.7$\pm$2.9} 
\newcommand{\TOITwoaRsLC}{34.2$^{+1.9}_{-4.6}$} 
\newcommand{\TOITwoi}{89.5$^{+0.4}_{-0.6}$} 
\newcommand{\TOITwob}{0.30$^{+0.27}_{-0.21}$} 
\newcommand{\TOITwoIns}{3.7$\pm$0.5} 
\newcommand{\TOITwoTeqUpper}{388} 
\newcommand{\TOITwoTeqHigh}{355} 
\newcommand{\TOITwoTeqLow}{274} 

\begin{document}

\title{TOI 122b and TOI 237b, two small warm planets orbiting inactive M dwarfs, found by \textit{TESS}}

\author[0000-0002-8961-0352]{William~C.~Waalkes}
\altaffiliation{NSF Graduate Research Fellow}
\affiliation{Department of Astrophysical \& Planetary Sciences, University of Colorado Boulder, 2000 Colorado Ave, Boulder, CO 80309, USA}

\author[0000-0002-3321-4924]{Zachory~K.~Berta-Thompson}
\affiliation{Department of Astrophysical \& Planetary Sciences, University of Colorado Boulder, 2000 Colorado Ave, Boulder, CO 80309, USA}

\author[0000-0001-6588-9574]{Karen~A.~Collins}
\affiliation{Center for Astrophysics \textbar \ Harvard \& Smithsonian, 60 Garden Street, Cambridge, MA 02138, USA}

\author[0000-0002-9464-8101]{Adina~D.~Feinstein}
\altaffiliation{NSF Graduate Research Fellow}
\affil{Department of Astronomy and Astrophysics, University of Chicago, 5640 S. Ellis Ave, Chicago, IL 60637, USA}

\author[0000-0003-2053-0749]{Benjamin~M.~Tofflemire}
\affiliation{Department of Astronomy, The University of Texas at Austin, Austin, TX 78712, USA}

\author[0000-0002-0149-1302]{B\'arbara Rojas-Ayala}
\affiliation{Instituto de Alta Investigaci\'on, Universidad de Tarapac\'a, Casilla 7D, Arica, Chile}

\author[0000-0003-2565-7909]{Michele~L.~Silverstein}
\altaffiliation{NASA Postdoctoral Program Fellow}
\affiliation{NASA Goddard Space Flight Center, Greenbelt, MD 20771, USA}
\affiliation{RECONS Institute, Chambersburg, PA 17201, USA}

\author[0000-0003-4150-841X]{Elisabeth Newton}
\affiliation{Department of Physics and Astronomy, Dartmouth College, Hanover NH 03755, USA}

\nocollaboration{}

\author[0000-0003-2058-6662]{George~R.~Ricker}
\affiliation{Department of Physics and Kavli Institute for Astrophysics and Space Research, Massachusetts Institute of Technology, Cambridge, MA 02139, USA}

\author[0000-0001-6763-6562]{Roland~Vanderspek}
\affiliation{Department of Physics and Kavli Institute for Astrophysics and Space Research, Massachusetts Institute of Technology, Cambridge, MA 02139, USA}

\author[0000-0001-9911-7388]{David~W.~Latham}
\affiliation{Center for Astrophysics \textbar \ Harvard \& Smithsonian, 60 Garden Street, Cambridge, MA 02138, USA}

\author[0000-0002-6892-6948]{S.~Seager}
\affiliation{Department of Physics and Kavli Institute for Astrophysics and Space Research, Massachusetts Institute of Technology, Cambridge, MA 02139, USA}
\affiliation{Department of Earth, Atmospheric and Planetary Sciences, Massachusetts Institute of Technology, Cambridge, MA 02139, USA}
\affiliation{Department of Aeronautics and Astronautics, MIT, 77 Massachusetts Avenue, Cambridge, MA 02139, USA}

\author[0000-0002-4265-047X]{Joshua~N.~Winn}
\affiliation{Department of Astrophysical Sciences, Princeton University, 4 Ivy Lane, Princeton, NJ 08544, USA}

\author[0000-0002-4715-9460]{Jon~M.~Jenkins}
\affiliation{NASA Ames Research Center, Moffett Field, CA, 94035, USA}

\author{Jessie Christiansen}
\affiliation{Caltech/IPAC-NASA Exoplanet Science Institute, 770 S. Wilson Avenue, Pasadena, CA 91106, USA}

\author{Robert~F.~Goeke}
\affiliation{Department of Physics and Kavli Institute for Astrophysics and Space Research, Massachusetts Institute of Technology, Cambridge, MA 02139, USA}

\author[0000-0001-8172-0453]{Alan~M.~Levine}
\affiliation{Department of Physics and Kavli Institute for Astrophysics and Space Research, Massachusetts Institute of Technology, Cambridge, MA 02139, USA}

\author[0000-0002-4047-4724]{H.~P.~Osborn}
\affiliation{Department of Physics and Kavli Institute for Astrophysics and Space Research, Massachusetts Institute of Technology, Cambridge, MA 02139, USA}
\affiliation{NCCR/PlanetS, Centre for Space \& Habitability, University of Bern, Bern, Switzerland}

\author[0000-0003-2519-3251]{S. A. Rinehart}
\affiliation{NASA HQ, Planetary Science Division
202-358-1884}

\author{Mark E. Rose}
\affiliation{NASA Ames Research Center, Moffett Field, CA, 94035, USA}

\author[0000-0002-8219-9505]{Eric B. Ting}
\affiliation{NASA Ames Research Center, Moffett Field, CA, 94035, USA}

\author[0000-0002-6778-7552]{Joseph~D.~Twicken}
\affiliation{NASA Ames Research Center, Moffett Field, CA, 94035, USA}
\affiliation{SETI Institute, Mountain View, CA 94043, USA}

\nocollaboration{}

\author[0000-0003-1464-9276]{Khalid Barkaoui} 
\affiliation{Astrobiology Research Unit, Universit\'e de Li\`ege, 19C All\`ee du 6 Ao\^ut, 4000 Li\`ege, Belgium} 
\affiliation{Oukaimeden Observatory, High Energy Physics and Astrophysics Laboratory, Cadi Ayyad University, Marrakech, Morocco}

\author{Jacob~L.~Bean}
\affil{Department of Astronomy and Astrophysics, University of
Chicago, 5640 S. Ellis Ave, Chicago, IL 60637, USA}

\author{C\'{e}sar Brice\~{n}o}
\affiliation{Cerro Tololo Interamerican Observatory /
NSF's National Optical-Infrared Astronomy Research Laboratory, Casilla 603, La Serena, Chile} 

\author[0000-0002-5741-3047]{David~ R.~Ciardi}
\affiliation{Caltech/IPAC-NASA Exoplanet Science Institute, 770 S. Wilson Avenue, Pasadena, CA 91106, USA}

\author[0000-0003-2781-3207]{Kevin I.\ Collins}
\affiliation{George Mason University, 4400 University Drive, Fairfax, VA 22030, USA}

\author[0000-0003-2239-0567]{Dennis Conti}
\affiliation{American Association of Variable Star Observers, 49 Bay State Rd, Cambridge, MA 02138, USA}

\author[0000-0002-4503-9705]{Tianjun Gan}
\affil{Department of Astronomy and Tsinghua Centre for Astrophysics, Tsinghua University, Beijing 100084, China}

\author[0000-0003-1462-7739]{Micha\"el Gillon} 
\affiliation{Astrobiology Research Unit, Universit\'e de Li\`ege, 19C All\`ee du 6 Ao\^ut, 4000 Li\`ege, Belgium}

\author{Giovanni Isopi}
\affiliation{Campo Catino Astronomical Observatory, Regione Lazio, Guarcino (FR) 03010, Italy}

\author{Emmanu\"el Jehin}
\affiliation{Space Sciences, Technologies and Astrophysics Research (STAR) Institute, Universit\'e de Li\`ege, 19C All\`ee du 6 Ao\^ut, 4000 Li\`ege, Belgium}

\author[0000-0002-4625-7333]{Eric L. N. Jensen}
\affiliation{Dept.\ of Physics \& Astronomy, Swarthmore College, Swarthmore PA 19081, USA}

\author[0000-0003-0497-2651]{John F.\ Kielkopf}
\affiliation{Department of Physics and Astronomy, University of Louisville, Louisville, KY 40292, USA}

\author{Nicholas Law}
\affiliation{Department of Physics and Astronomy, The University of North Carolina at Chapel Hill, Chapel Hill, NC 27599-3255, USA}

\author{Franco Mallia}
\affiliation{Campo Catino Astronomical Observatory, Regione Lazio, Guarcino (FR) 03010, Italy}

\author[0000-0003-3654-1602]{Andrew W. Mann}
\affiliation{Department of Physics and Astronomy, The University of North Carolina at Chapel Hill, Chapel Hill, NC 27599-3255, USA}

\author[0000-0001-7516-8308]{Benjamin~T.~Montet}
\altaffiliation{Sagan Fellow}
\affiliation{School of Physics, University of New South Wales, Sydney NSW 2052, Australia}

\author[0000-0003-1572-7707]{Francisco J. Pozuelos}
\affiliation{Space Sciences, Technologies and Astrophysics Research (STAR) Institute, Universit\'e de Li\`ege, 19C All\`ee du 6 Ao\^ut, 4000 Li\`ege, Belgium}
\affiliation{Astrobiology Research Unit, Universit\'e de Li\`ege, 19C All\`ee du 6 Ao\^ut, 4000 Li\`ege, Belgium}

\author{Howard Relles}
\affiliation{Center for Astrophysics \textbar \ Harvard \& Smithsonian, 60 Garden Street, Cambridge, MA 02138, USA}

\author[0000-0002-2990-7613]{Jessica E. Libby-Roberts}
\affiliation{Department of Astrophysical \& Planetary Sciences, University of Colorado Boulder, 2000 Colorado Ave, Boulder, CO 80309, USA}

\author{Carl Ziegler}
\affiliation{Dunlap Institute for Astronomy and Astrophysics, University of Toronto, 50 St. George Street, Toronto, Ontario M5S 3H4, Canada}

\begin{abstract}
We report the discovery and validation of TOI 122b and TOI 237b, two warm planets transiting inactive M dwarfs observed by \textit{TESS}. Our analysis shows TOI 122b has a radius of \TOIOneRp~R$_{\Earth}$ and receives \TOIOneIns$\times$ Earth's bolometric insolation, and TOI 237b has a radius of \TOITwoRp~R$_{\Earth}$ and receives \TOITwoIns$\times$ Earth insolation, straddling the 6.7$\times$ Earth insolation that Mercury receives from the sun. This makes these two of the cooler planets yet discovered by \textit{TESS}, even on their 5.08-day and 5.43-day orbits. Together, they span the small-planet radius valley, providing useful laboratories for exploring volatile evolution around M dwarfs. Their relatively nearby distances (\TOIOneDist~pc and \TOITwoDist~pc, respectively) make them potentially feasible targets for future radial velocity follow-up and atmospheric characterization, although such observations may require substantial investments of time on large telescopes. \end{abstract}

\section{Introduction}

The Transiting Exoplanet Survey Satellite \citep[\textit{TESS},][]{Ricker2015} follows the 8 year missions of \textit{Kepler} \citep{Borucki2010} and \textit{K2} \citep{Howell2014}, which discovered thousands of planets. While \textit{Kepler} typically found planets orbiting faint and distant stars, \textit{TESS} is examining the brightest and nearest stars for evidence of exoplanet transits. Over the course of its 2-year primary mission, \textit{TESS} has surveyed 85\% of the sky, looking at over 200,000 nearby stars with a 2-minute cadence and many more stars with the 30-minute full frame images (FFIs). \textit{TESS} is expected to find up to 4500 planets, 500-1200 planets orbiting M dwarfs, and about 50 planets within 50 pc \citep[see][]{Sullivan2015,Barclay2018,Ballard2019}.

M dwarfs are interesting targets for transiting exoplanet studies as they provide the best opportunity for finding temperate terrestrial planets \citep{Nutzman2008,Blake2008}. All main sequence stars less massive than 0.6~M$_{\sun}$ fall into the M dwarf category, and they are the most numerous stellar type in the universe \citep[e.g.,][]{Chabrier2000}. These stars are very cool (2000~\rm{K}$<$\teff$<$4000\rm{K}) and very small, so cool planets have shorter periods, higher transit probabilities and deeper transits than they would around larger stars.

M dwarfs tend to host terrestrial exoplanets more often than gas giants \citep{Mulders2015,Bowler2015}, and these terrestrial planets can more readily be found at lower insolations given the low luminosities of M dwarfs. Finally, M dwarfs have such long lifetimes that not a single M dwarf ever formed has yet evolved off the main sequence \citep{Laughlin1997}, making these stellar systems interesting laboratories for very long timescale planetary evolution. For a comprehensive review of M dwarfs as exoplanet host stars, see \citet{Shields2019}. While the habitability of planets around M dwarfs remains an open question, the low insolations of M dwarf planets on short periods creates opportunities for statistically studying the presence and evolution of planetary atmospheres.

\begin{figure*}[t!]
\centering
\includegraphics[width=0.99\textwidth]{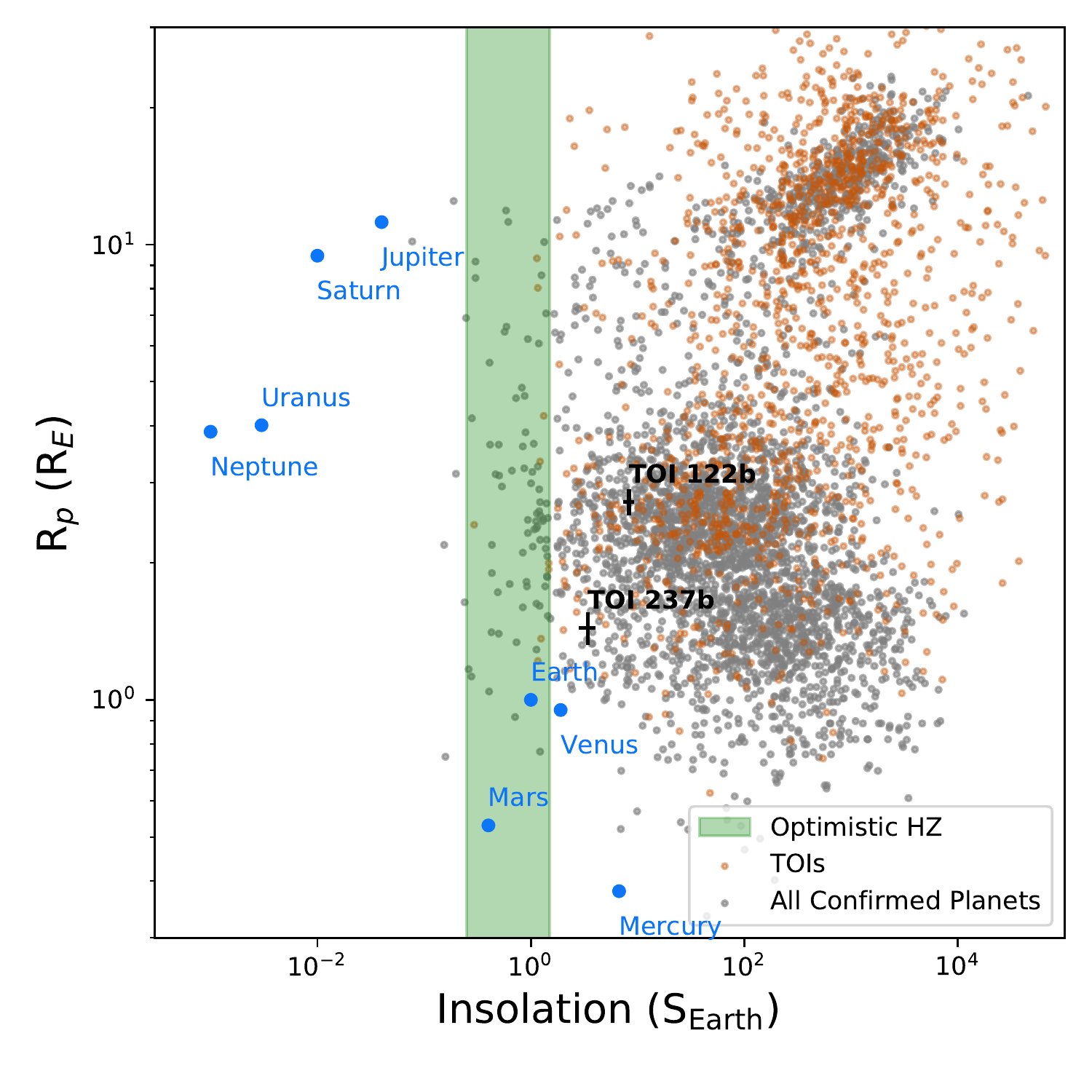}
\caption{All confirmed exoplanets and current \textit{TESS} Objects of Interest (TOIs) (as of February 2020) with current values for R$_{\rm{p}}$ (R$_{\Earth}$) and S (S$_{\Earth}$). Orange points are the TOIs (validated and unvalidated), while the gray points are all confirmed exoplanets (as of March 2020). Highlighted in green is the ``recent Venus-early Mars" habitable zone covering 0.25-1.5 S$_{\Earth}$ \citep[e.g.,][]{Kopparapu2019}, in which a few systems fall. This optimistic habitable zone is likely shifted to lower insolations for M dwarfs given more recent studies of energy budgets and albedos for M dwarf planets \citep{Shields2019}. \label{Rp_insol}}
\end{figure*}

The first year of \textit{TESS} yielded several small exoplanets orbiting M dwarfs such as LHS 3844b \citep{Vanderspek2019}, the L 98-59 system \citep{Kostov2019,Cloutier2019}, the TOI 270 system \citep{Guenther2019}, the Gl 357 system \citep{Luque2019}, LTT 1445Ab \citep{Winters2019}, the LP 791-18 system \citep{Crossfield2019}, and L 168-9b \citep{Astudillo-Defru2020}. The two planets we present in this paper are challenging for precise RV mass measurements, but both are smaller than 3~R$_\Earth$ and their low insolations and short periods (see Fig. \ref{Rp_insol}) position them as interesting candidates for atmospheric follow-up. They may have retained their atmospheres despite being too hot to be considered habitable and may help us understand atmospheric evolution and the diversity of atmospheres of small planets.

The rest of the paper is organized as follows. In \S 2 we describe the \textit{TESS} observations, the photometric and spectroscopic follow-up data we gathered, and arguments against these planets being false positives. In \S 3 we describe the results of stellar parameter estimation and transit light curve fitting, and in \S 4 we discuss the results and their implications for future work.

\begin{table*}
\begin{center}
\begin{tabular}{cccccc}
\toprule
    Date & Observatory & Filter & Exposure Time (s) & Aperture Radius (\arcsec) & Transit Midpoint (BJD TDB)\\
    \midrule
    \textbf{TOI 122b}\\
    2018-09-18 & SSO iTelescope & Clear & 120 & 4.8 & (Egress Only) \\
    \multirow{ 2}{*}{2018-09-18} & LCO SSO (1m) & r' & 180 & 4.28 & \multirow{ 2}{*}{2458379.901563$^{+0.0.001239}_{-0.001189}$} \\
     & LCO SSO (1m) & i' & 30 & 3.89 &  \\
    2018-10-18 & LCO SAAO (1m) & I & 42 & 5.45 & (Too Noisy) \\
    \multirow{ 2}{*}{2018-11-02} & TRAPPIST South (0.6m) & I+z' & 60 & 5.2 & \multirow{ 2}{*}{2458425.602564$^{+0.000630}_{-0.000633}$} \\
     & LCO CTIO (1m) & I & 42 & 4.67 &  \\
    2019-07-10 & LCO SAAO (1m) & I & 50 & 5.06 & 2458674.427546$^{+0.000773}_{-0.000751}$ \\
    2019-07-15 & LCO SAAO (1m) & I & 50 & 3.50 & (Too Noisy) \\
    \multirow{ 2}{*}{2019-07-25} & LCO CTIO (1m) & g' & 240 & 4.67 & \multirow{ 2}{*}{2458689.657695$^{+0.001223}_{-0.001122}$} \\
     & LCO CTIO (1m) & g' & 240 & 3.89 &  \\
    2019-08-04 & LCO CTIO (1m) & V & 240 & 5.45 & 2458699.817618$^{+0.001866}_{-0.001965}$ \\
    \midrule
    \textbf{TOI 237b } \\
    2018-12-16 & LCO SAAO (1m) & i' & 65 & 4.67 & (Bad Ephemeris) \\
    2019-05-07 & LCO SAAO (1m) & i' & 100 & 3.89 & (Bad Ephemeris) \\
    2019-06-02 & TRAPPIST South (0.6m) & I+z'  & 60 & 5.2 & 2458637.922471$^{+0.001419}_{-0.001352}$ \\
    2019-06-14 & LCO CTIO (1m) & I & 60 & 6.22 & 2458648.797058$^{+0.001854}_{-0.001739}$ \\
    2019-06-19 & LCO CTIO (1m) & I & 60 & 8.56 & (Too Noisy) \\
    \multirow{ 2}{*}{2019-08-02} & LCO CTIO (1m) & I & 75 & 5.45 & \multirow{ 2}{*}{2458697.7197997$^{+0.000796}_{-0.000815}$} \\
     & LCO CTIO (1m) & g' & 300 & 4.67 &  \\
    2019-08-13 & LCO SAAO (1m) & I & 70 & 4.67 & 2458708.592274$^{+0.001521}_{-0.001178}$ \\
    2019-09-03 & LCO SAAO (1m) & I & 70 & 5.06 & 2458730.341868$^{+0.001127}_{-0.001559}$ \\
\bottomrule
\end{tabular}
\end{center}
\caption{Ground-based follow-up observations of the two planets, with mid-transit times (if a transit is detected), exposure times, and filters. For data sets in which a transit is not detected, this could be due to the transit being missed entirely, or the transit being obscured by noise. LCO is the Las Cumbres Observatory which includes SAAO, the South African Astronomical Observatory, CTIO, the Cerro-Telolo Interamerican Observatory, and SSO, the telescopes at the Siding Spring Observatory. SSO iTelescope is the Siding Spring Observatory iTelescope, which is not part of the LCO network. Observations from this site unfortunately missed most of the transit so we do not include these data in our analysis. We report mid-transit times based on the joint modeling described in the text. \label{tab:obs}}
\end{table*}

\section{Data}

\subsection{TESS Photometry}

\begin{figure*}[t!]
\centering
\subfloat[]{\includegraphics[width=0.49\textwidth]{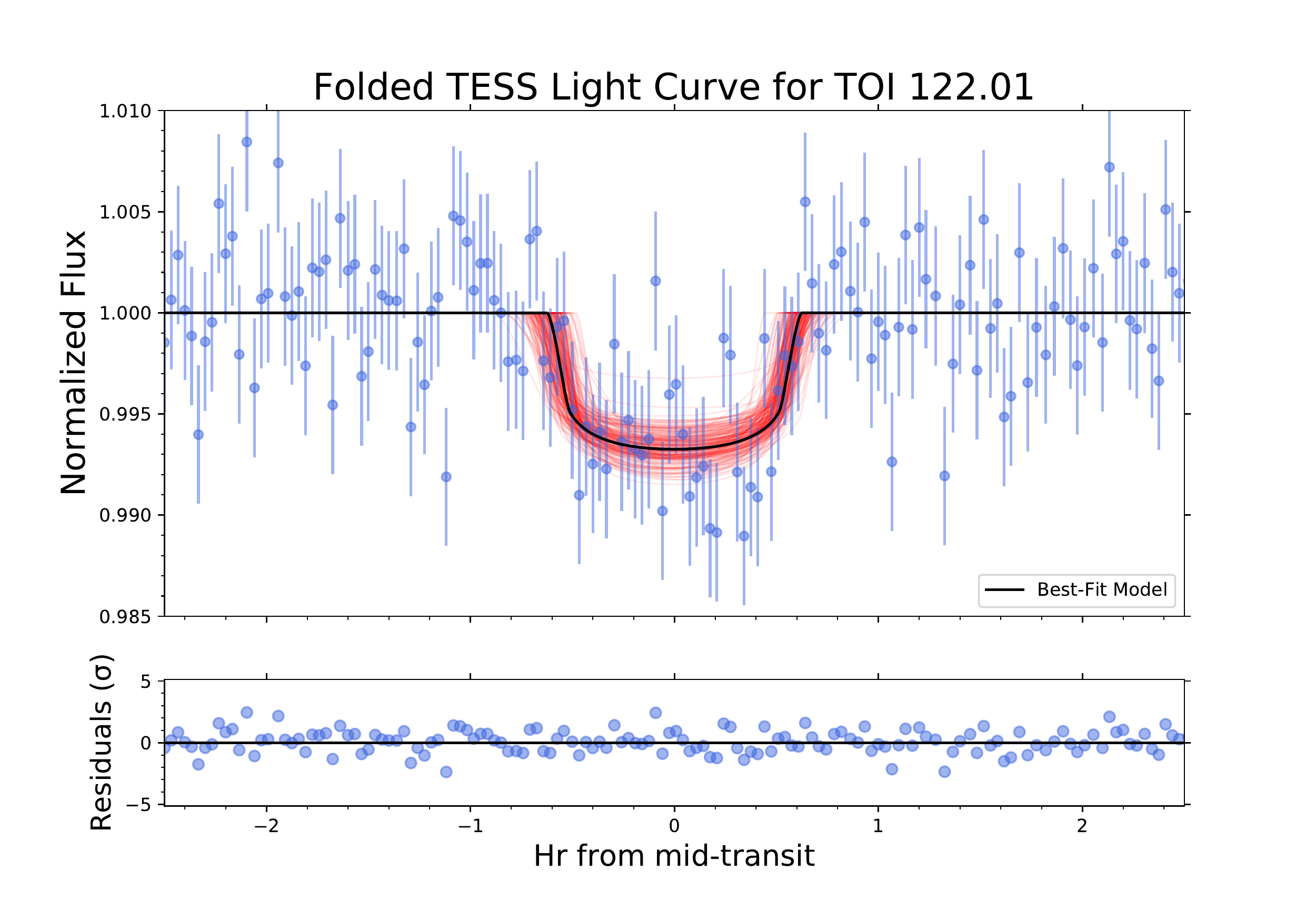}}
\subfloat[]{\includegraphics[width=0.49\textwidth]{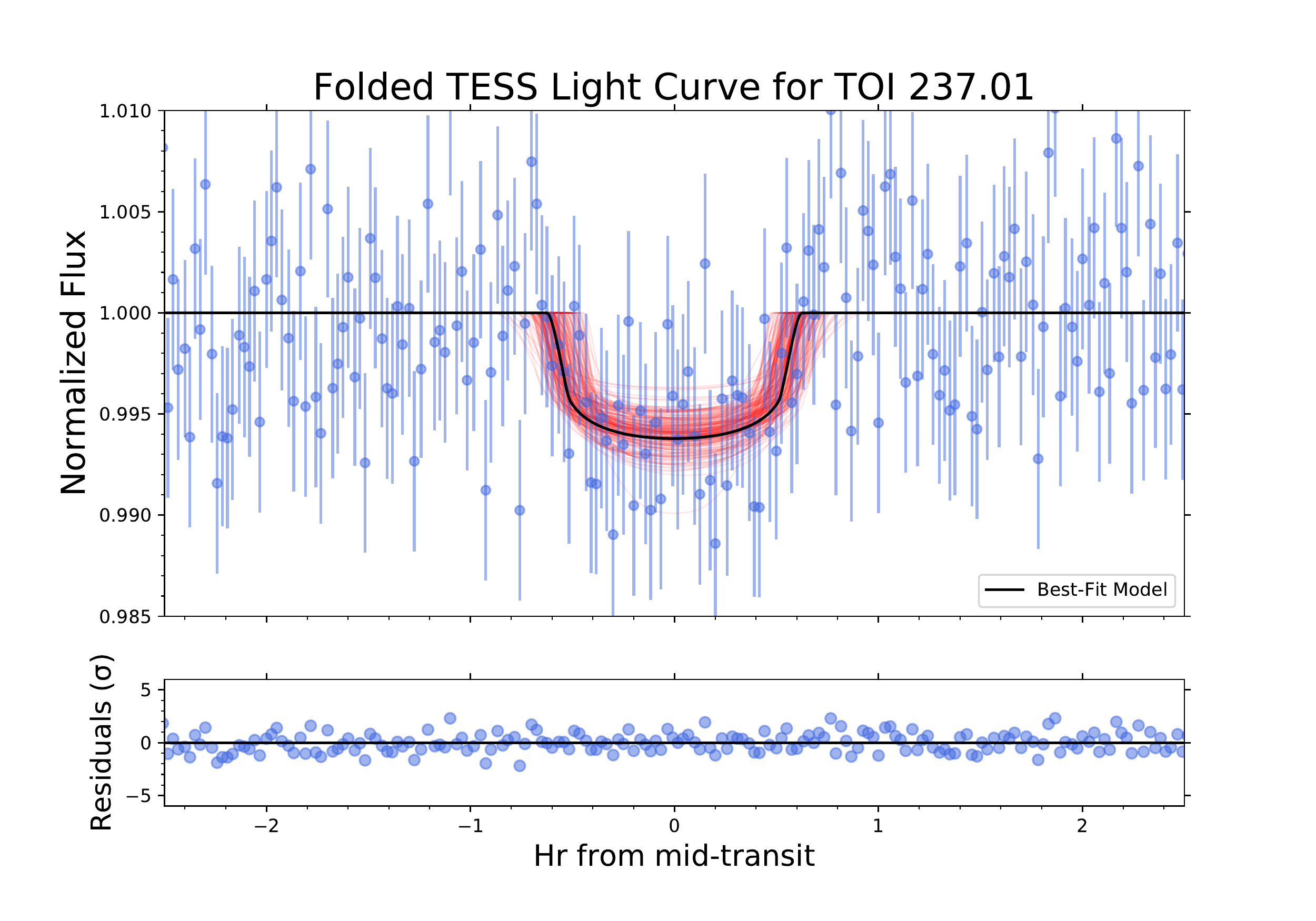}}
\caption{\textit{TESS} light curves, phase-folded across a full 27-day sector to the periods refined in this work. We model these light curves with a 3-parameter MCMC that explores values for transit depth, inclination, and the scaled semi-major axis. The best fit model (50$^{th}$ percentile values) is the black line, and red lines are random samples drawn from the posterior distributions. The posteriors from the \textit{TESS} light curves are consistent with the posteriors for the follow-up observations, with larger uncertainties. The follow-up observations have allowed us to constrain the transit parameters effectively.\label{TESSLightcurves}}
\end{figure*}

\textit{TESS} has four 24\degree$\times$24\degree~field of view cameras, each with four 2k$\times$2k CCDs. The \textit{TESS} bandpass is 600-1000~nm, and the pixel scale is 21 arcseconds \citep{Ricker2015}. For our analysis of the \textit{TESS} light curves (Fig. \ref{TESSLightcurves}), we accessed the \textit{TESS} data using \texttt{lightkurve} \citep{lightkurve} and downloaded the Science Processing Operations Center \citep[SPOC,][]{jenkinsSPOC2016} Presearch Data Conditioning Simple Aperture Photometry (PDCSAP) flux light curves \citep{Stumpe2012,2012PASP..124.1000S,2014PASP..126..100S}. The light curves shown in Figure \ref{TESSLightcurves} are 2-minute cadence data phase-folded to the orbital periods we refined in this work.

\textbf{\textit{TESS} Object of Interest (TOI) 122b} (TIC 231702397) was observed in Sector 1 of \textit{TESS} from 2018 July 25 to 2018 August 22 with CCD 1 of Camera 2. Four transits were observed with a 5.1 day period and a 6 ppt depth. The SPOC \citep{Jenkins2016} pipeline flagged the light curve as a planet candidate and it was submitted to the MIT TOI alerts page\footnote{\href{https://tess.mit.edu/toi-releases/}{https://tess.mit.edu/toi-releases/}} (Guerrero et al., submitted), where we accessed the preliminary SPOC data validation transit parameters \citep{Twicken:DVdiagnostics2018,Li:DVmodelFit2019} and scheduled follow-up observations with ground based observatories. Preliminary parameters indicated that the stellar host was an M dwarf, implying the orbiter was super-Earth or sub-Neptune in size.

\textbf{TOI 237b} (TIC 305048087) was observed in Sector 2 of \textit{TESS} from 2018 August 22 to 2018 September 20 with CCD 1 of Camera 1. Five transits were observed with a 5.4 day period and a 6 ppt depth. The SPOC pipeline flagged the light curve as a planet candidate and it was submitted to the MIT TOI alerts page, where we accessed the preliminary transit parameters and scheduled follow-up observations with ground based observatories. Preliminary parameters indicated that the stellar host was an M dwarf, implying the orbiter was also super-Earth in size.

\subsection{Ground-Based Photometry}

\begin{figure*}[t!]
\centering
\subfloat[]{\includegraphics[width=0.4\textwidth]{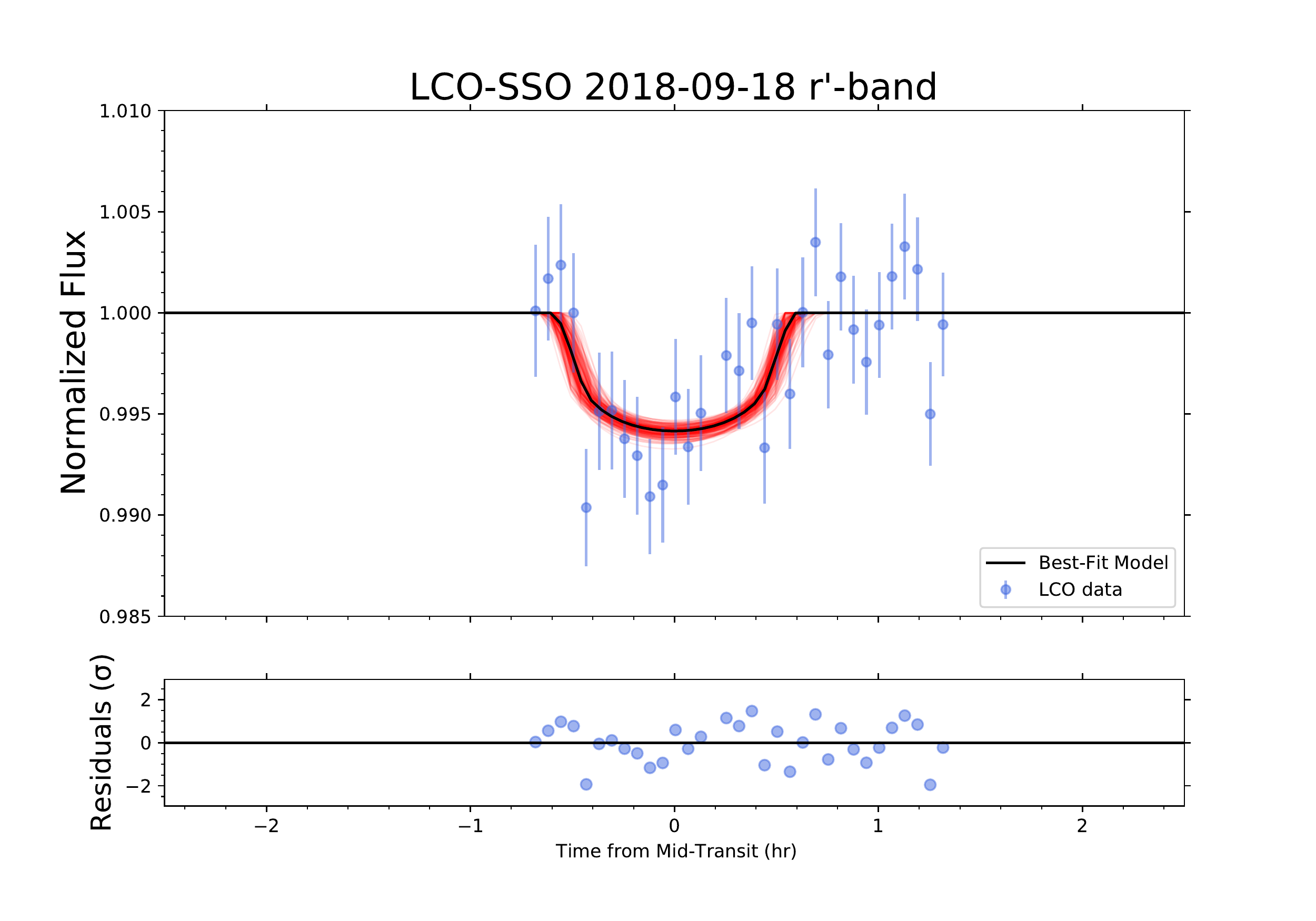}}
\subfloat[]{\includegraphics[width=0.4\textwidth]{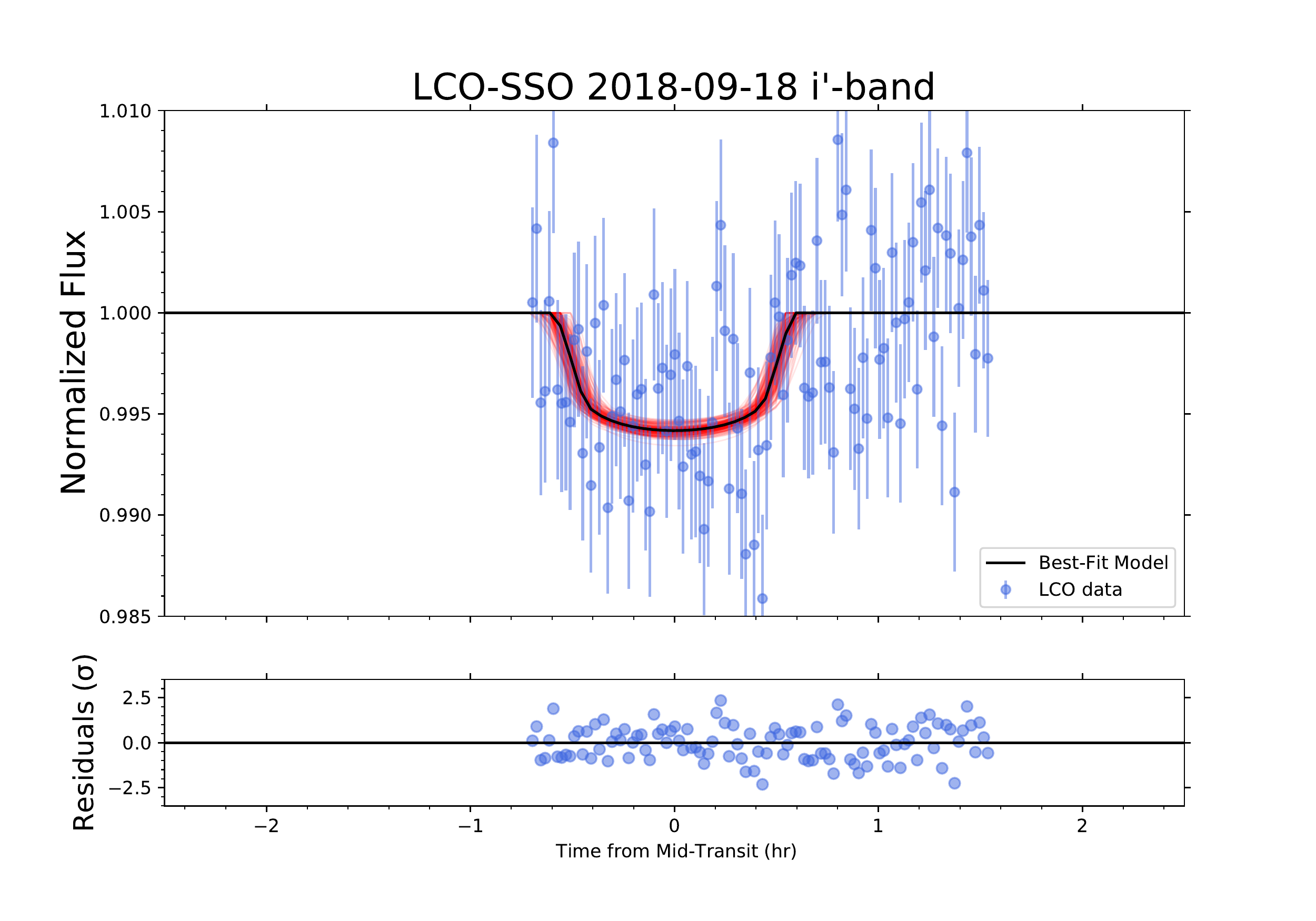}}\\
\subfloat[]{\includegraphics[width=0.4\textwidth]{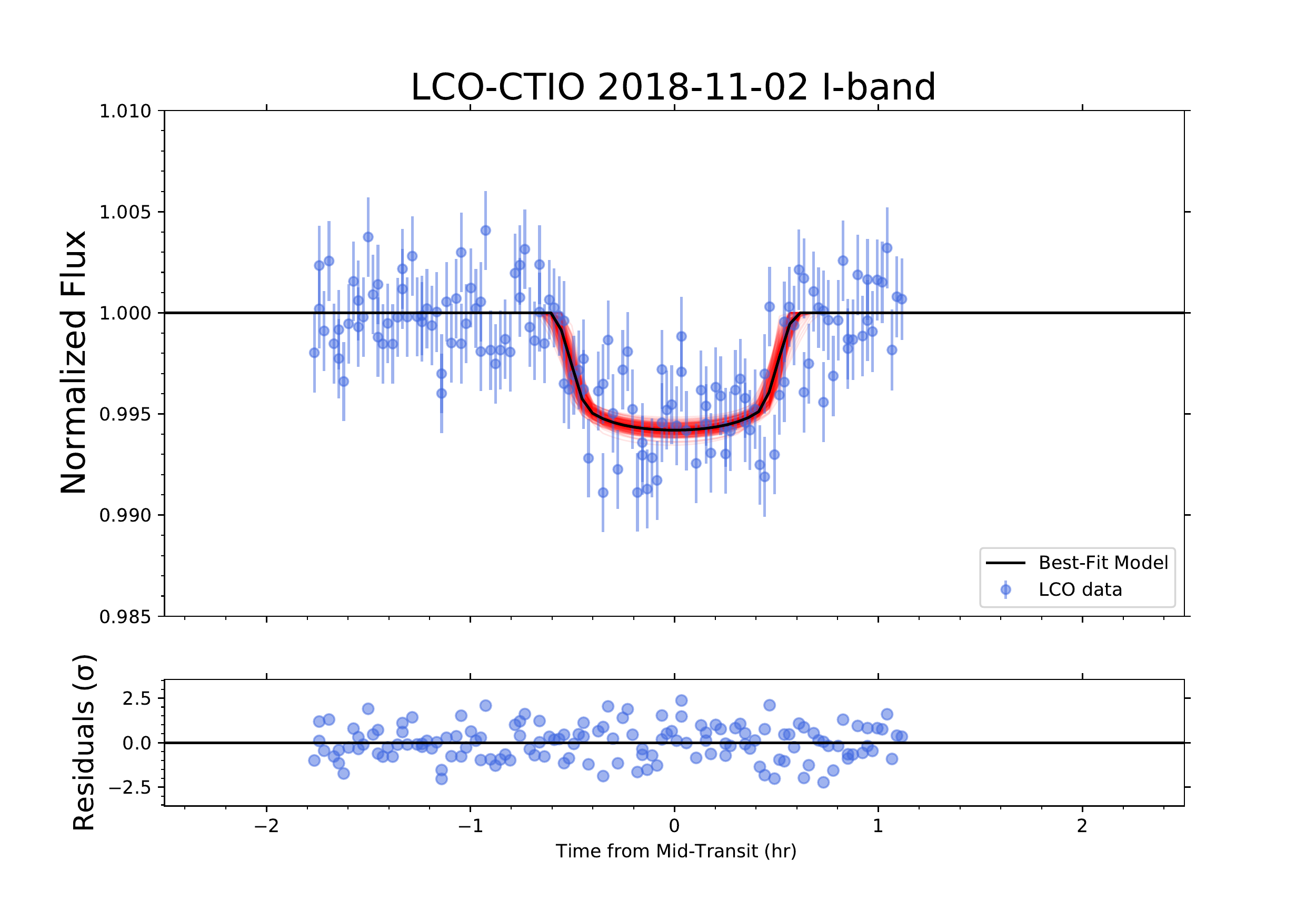}}
\subfloat[]{\includegraphics[width=0.4\textwidth]{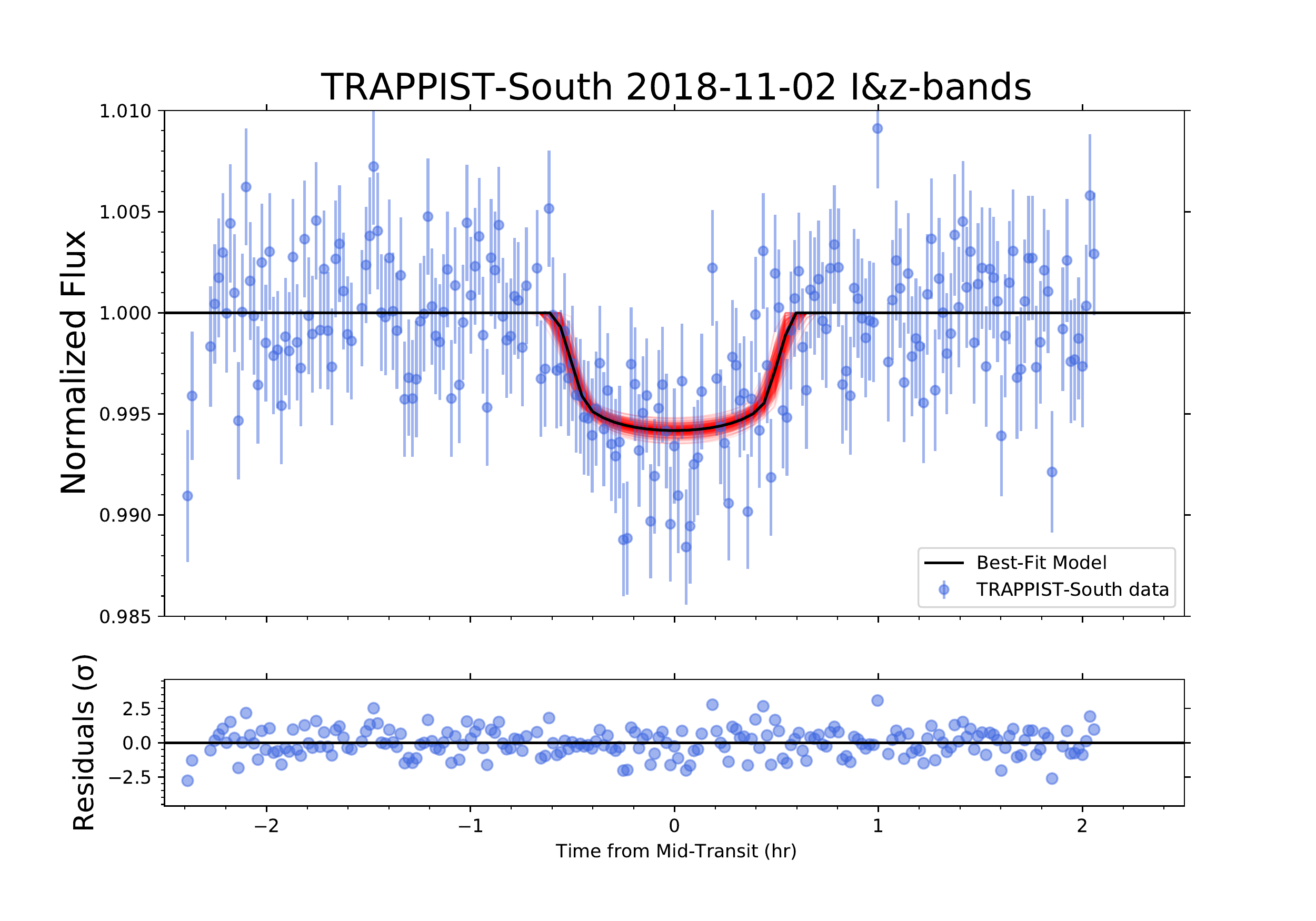}}\\
\subfloat[]{\includegraphics[width=0.4\textwidth]{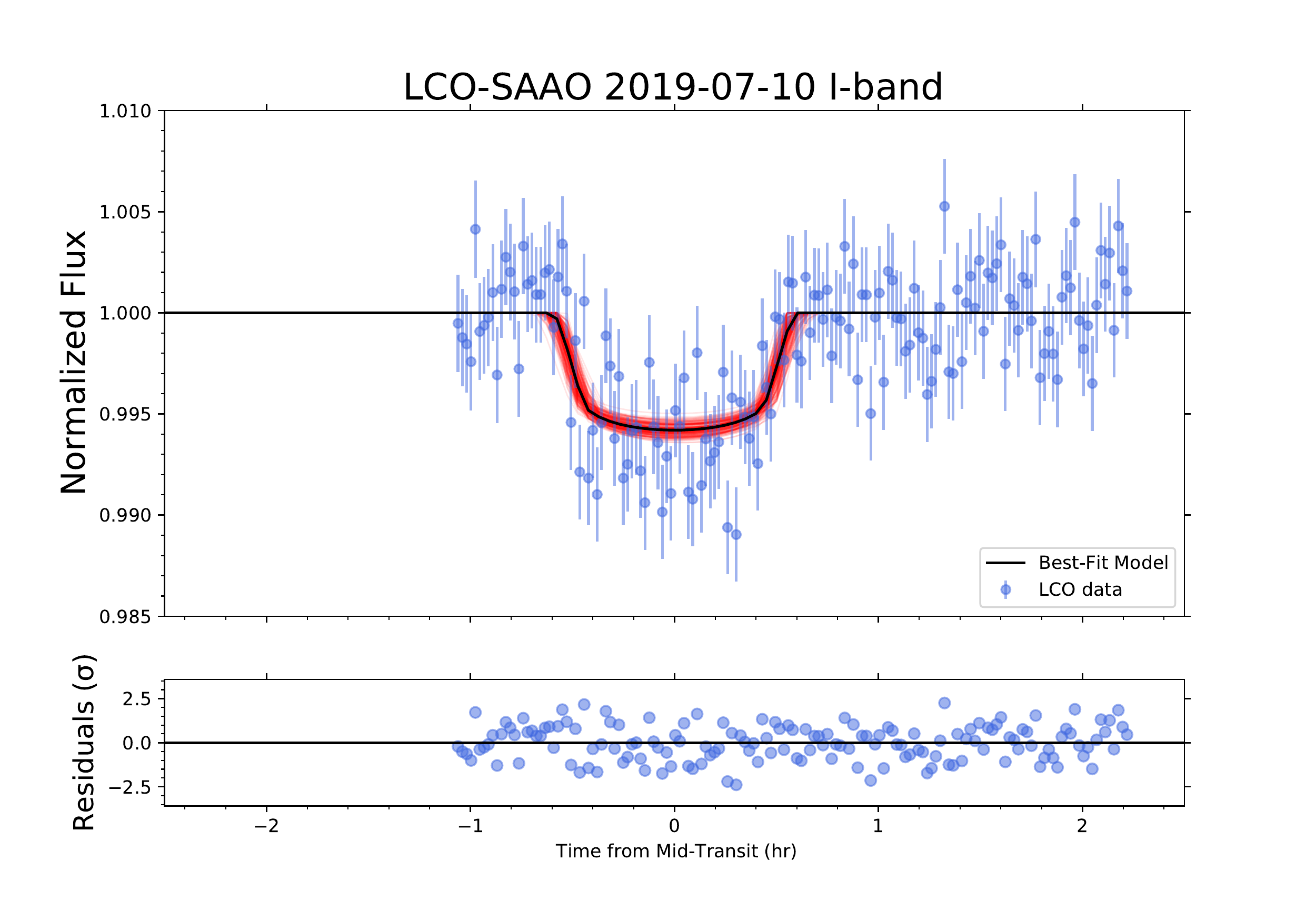}}
\subfloat[]{\includegraphics[width=0.4\textwidth]{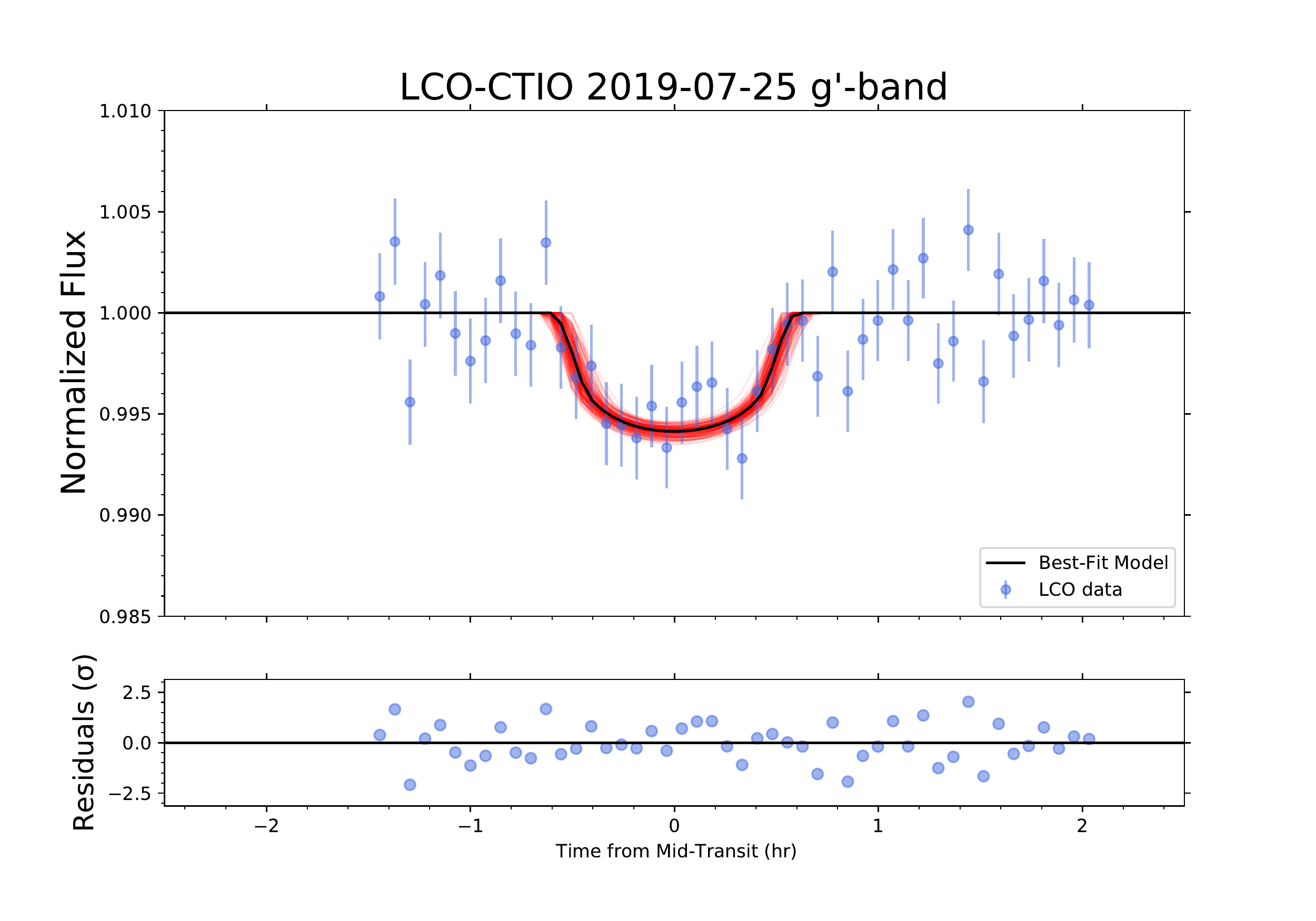}}\\
\subfloat[]{\includegraphics[width=0.4\textwidth]{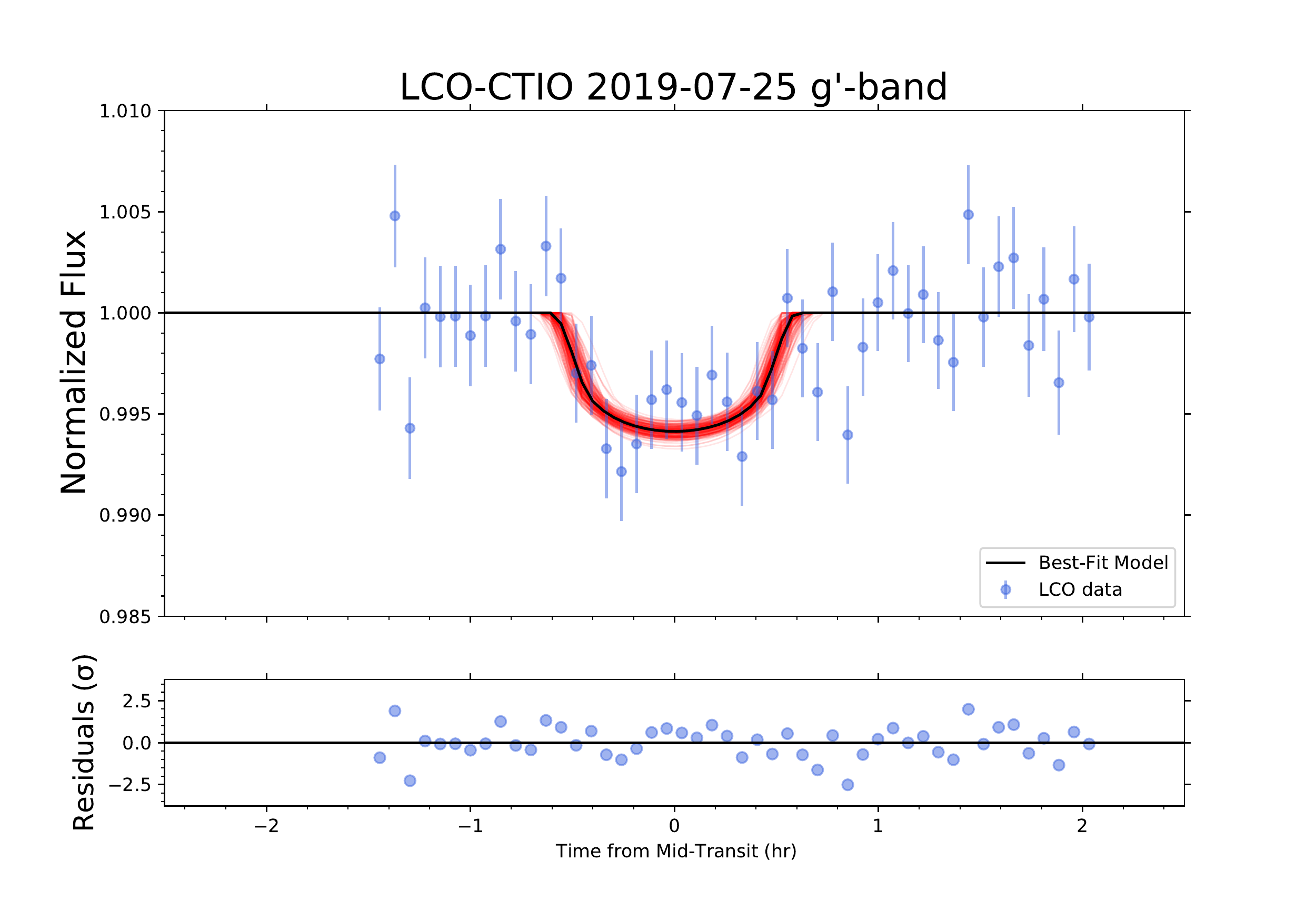}}
\subfloat[]{\includegraphics[width=0.4\textwidth]{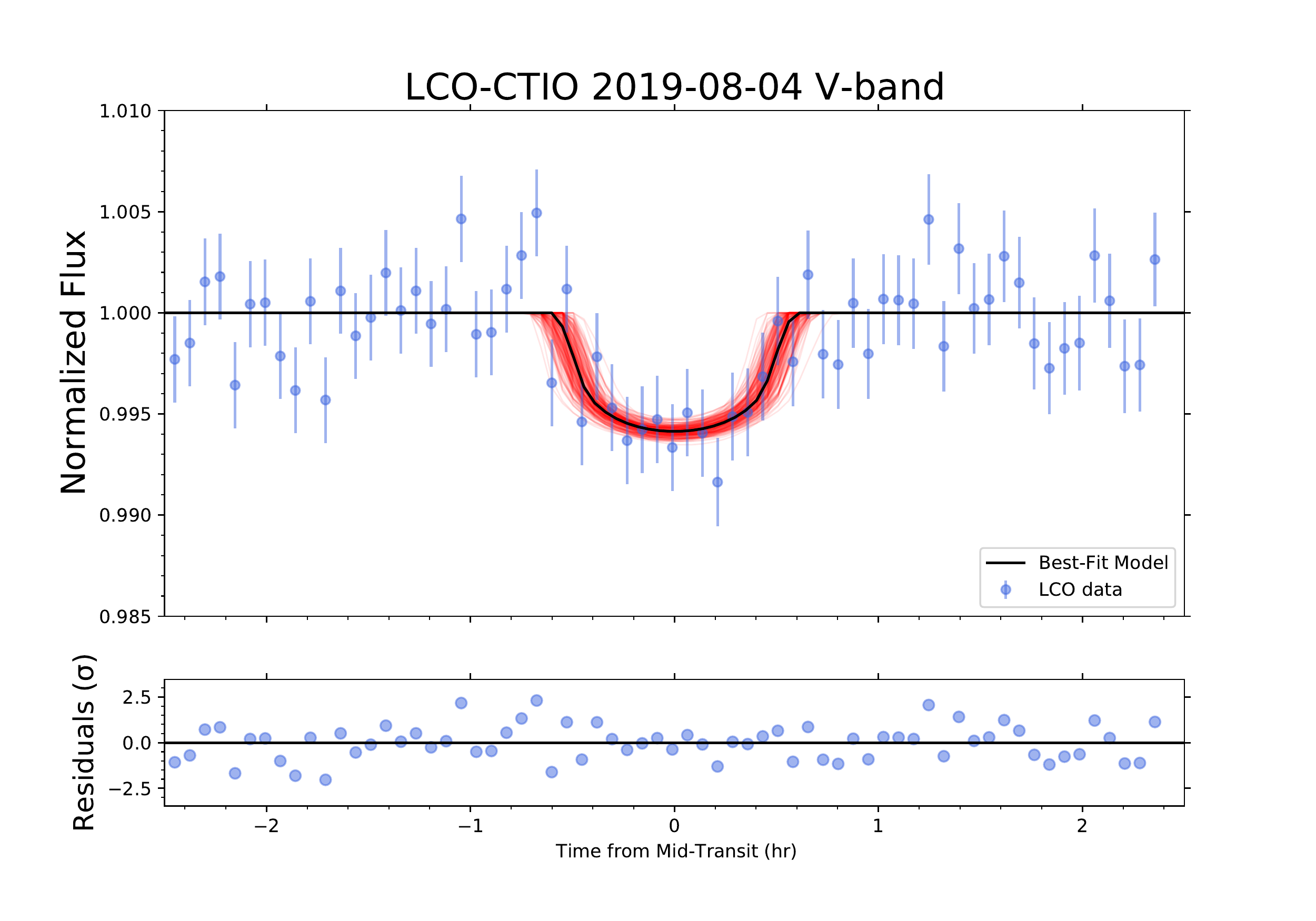}}
\caption{Light curves for all eight of the viable follow-up transits of TOI 122b. Best fit MCMC models are in black with 200 random samples plotted in red. Requiring that the transit depth, semi-major axis, and inclination were identical between visits led to a consistent model that fit all the transits.\label{TOI122_Lightcurves}}
\end{figure*}

\begin{figure*}[t!]
\centering
\subfloat[]{\includegraphics[width=0.48\textwidth]{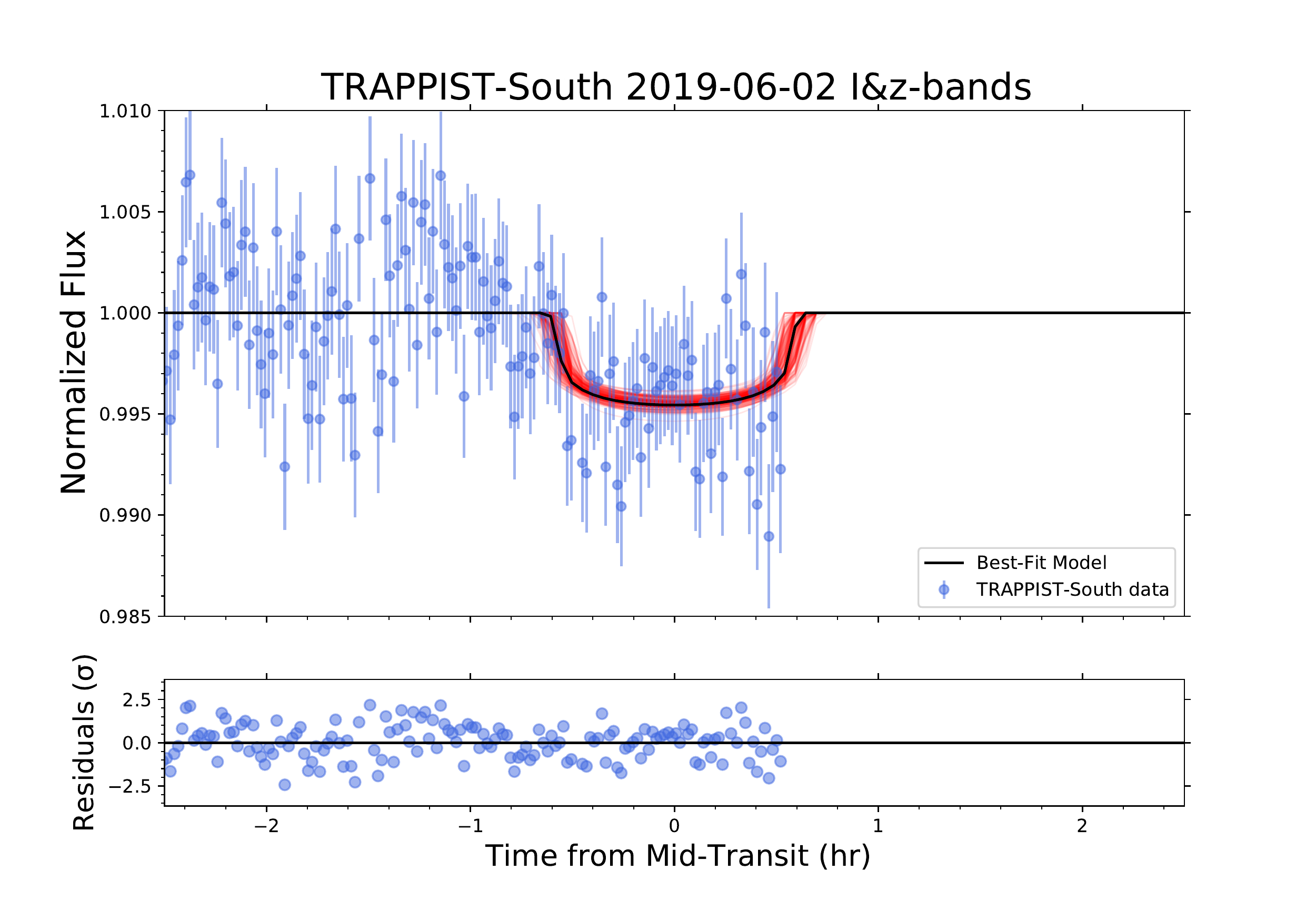}}
\subfloat[]{\includegraphics[width=0.48\textwidth]{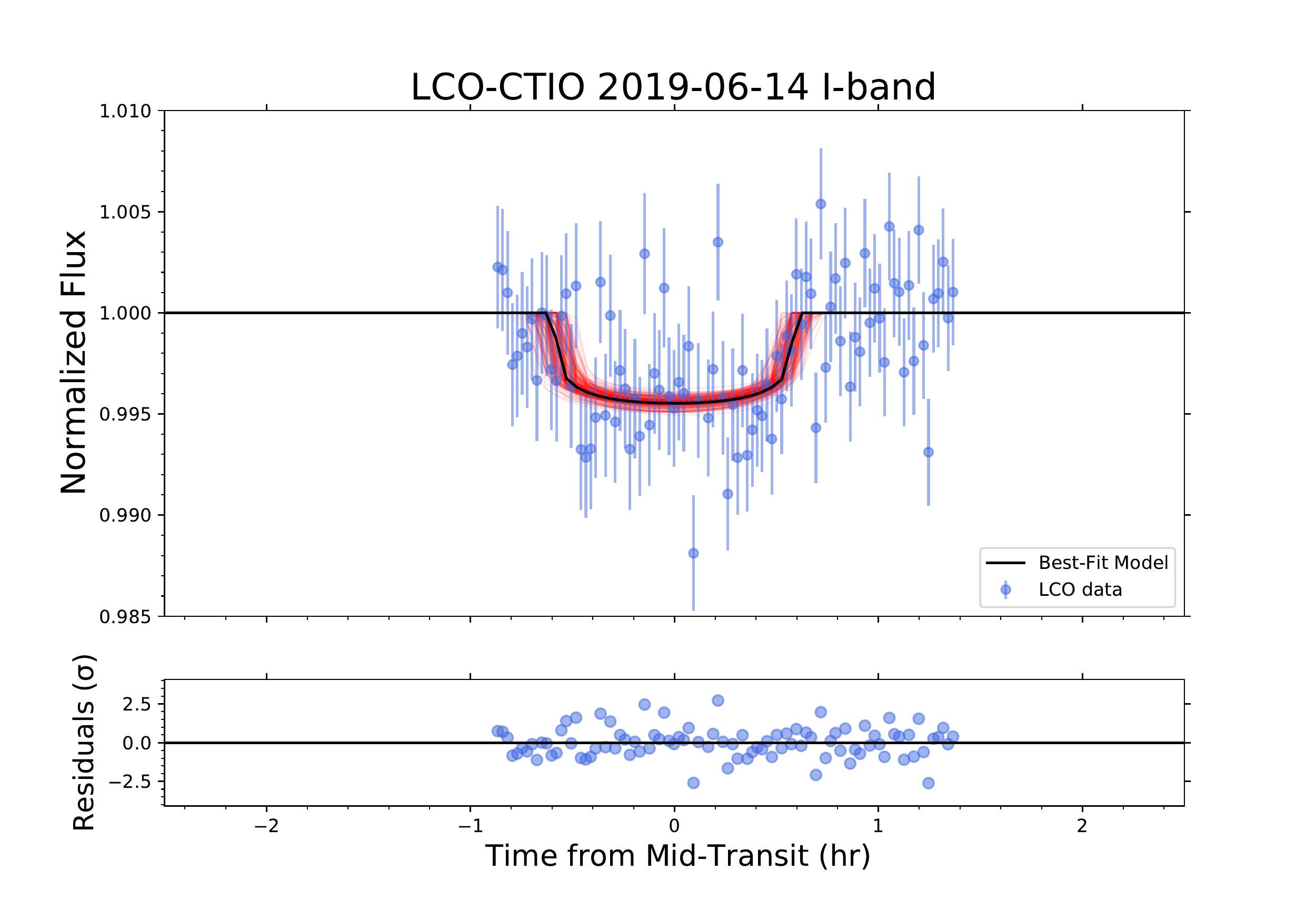}}\\
\subfloat[]{\includegraphics[width=0.48\textwidth]{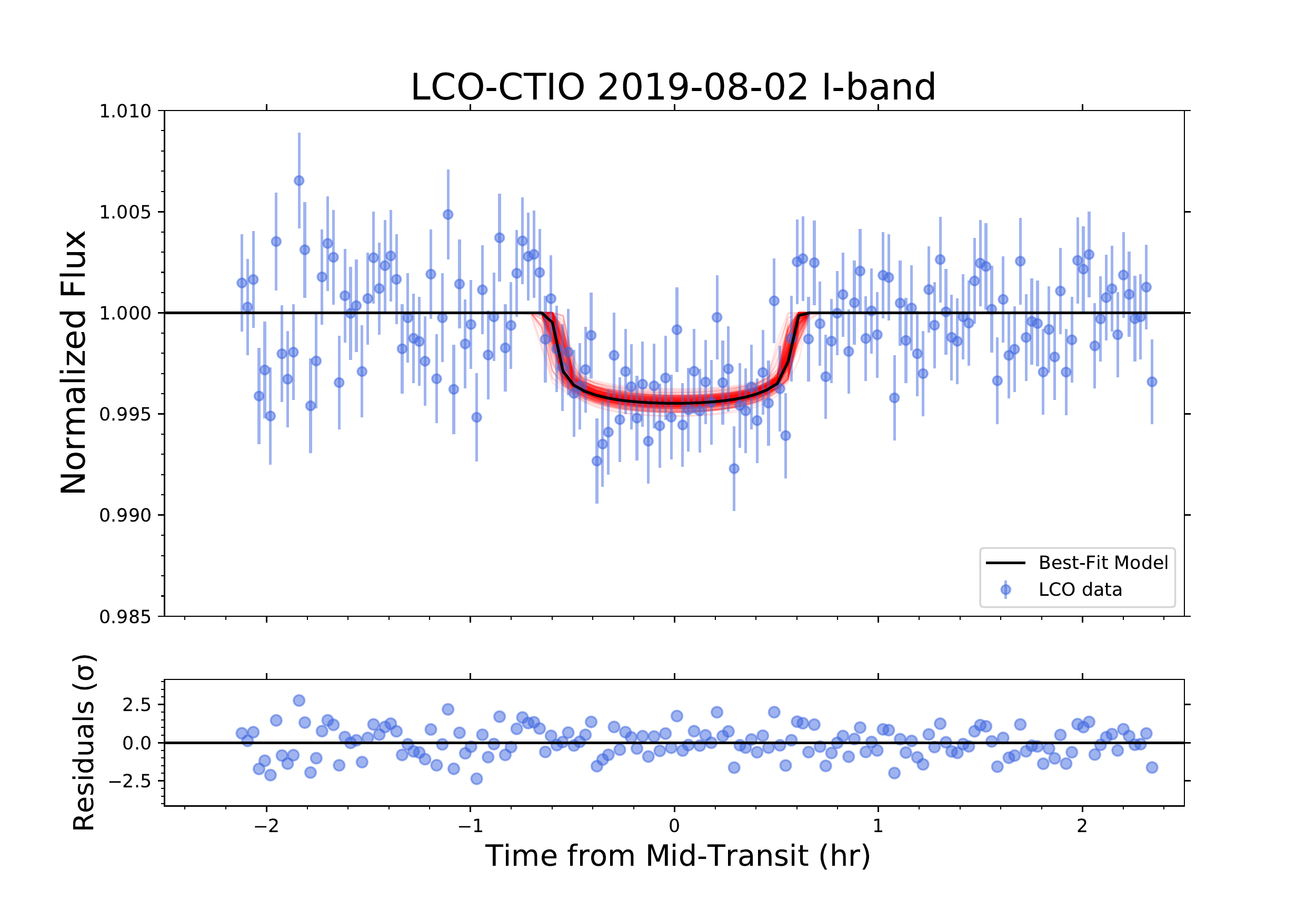}}
\subfloat[]{\includegraphics[width=0.48\textwidth]{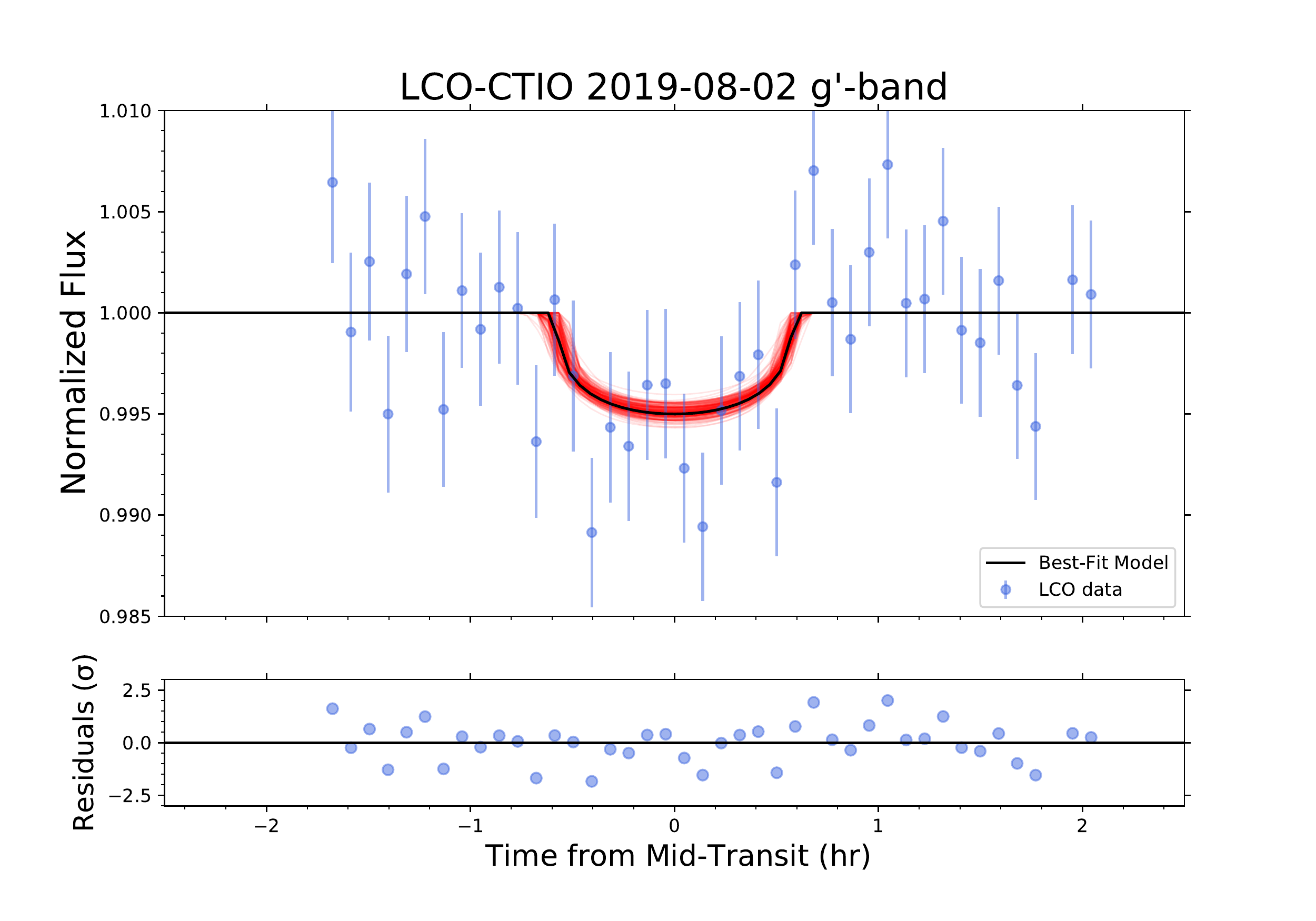}}\\
\subfloat[]{\includegraphics[width=0.48\textwidth]{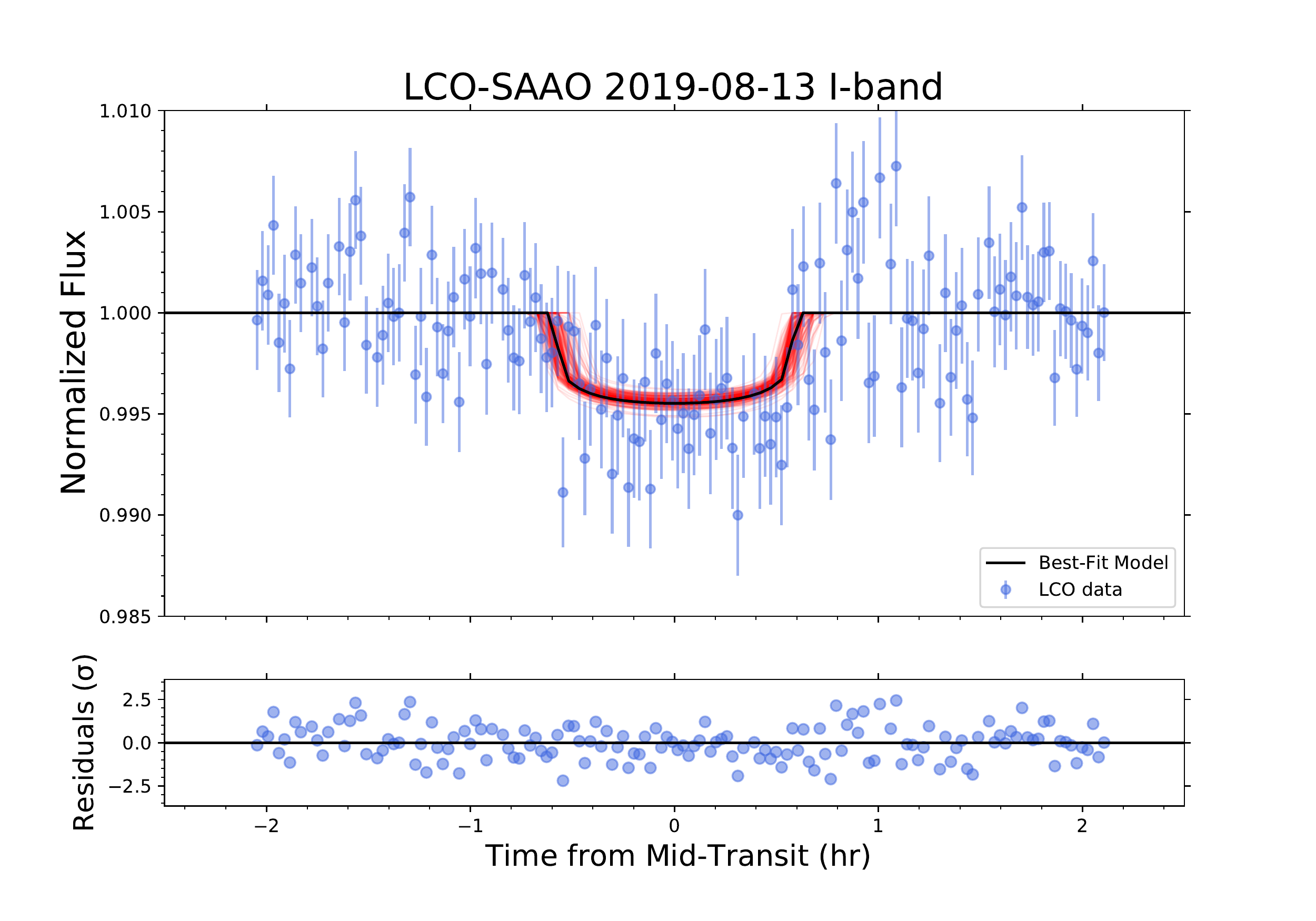}}
\subfloat[]{\includegraphics[width=0.48\textwidth]{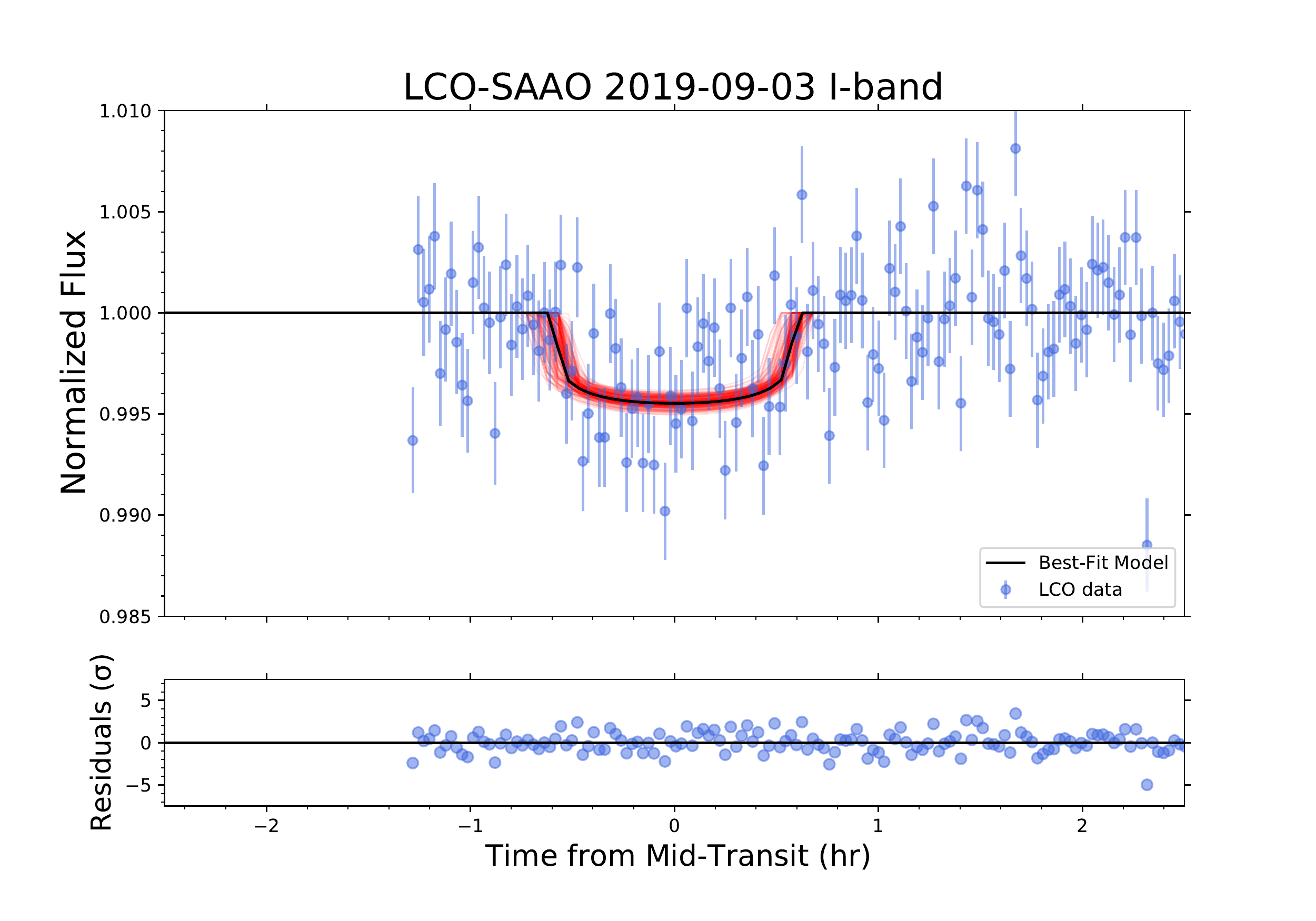}}
\caption{Light curves for ground-based follow-up transits of TOI 237b. Best fit MCMC models are in black with 200 random samples plotted in red. Requiring that the transit depth, semi-major axis, and inclination were consistent between visits led to a final model that fit all the transits.\label{TOI237_Lightcurves}
}
\end{figure*}

The follow-up observations are summarized in Table \ref{tab:obs}. Both systems were observed extensively as part of the \textit{TESS} Follow-up Observing Program Sub-Group 1 (TFOP SG1) photometric campaign. Ground-based observations span several months for both targets, from observatories around the globe. For both TOI 122 and TOI 237, we used the TESS Transit Finder tool, which is a customized version of the Tapir software package \citep{jensen2013}, to schedule the photometric time-series observations. Ground-based light curves used in the analysis are shown in Figures \ref{TOI122_Lightcurves} and \ref{TOI237_Lightcurves}.

\textbf{LCO Photometry}

Most photometric data were taken at Las Cumbres Observatory sites via the Las Cumbres Observatory Global Telescope (LCOGT) network \citep{Brown2013}. These observations were done with 1-m telescopes equipped with \textit{Sinistro} cameras which have a plate scale of 0.389 arcseconds and a FOV of $26.4\arcmin\times26.4\arcmin$. Filters and photometric aperture radii vary between observations and are provided in Table \ref{tab:obs}. Additional information and the full datasets can be found on ExoFOP-TESS\footnote{\href{https://exofop.ipac.caltech.edu/tess/}{https://exofop.ipac.caltech.edu/tess/}}.

LCOGT data are reduced via a standard reduction pipeline \citep[``BANZAI",][]{BANZAI} which performs bias and dark subtractions, flat field correction, bad pixel masking, astrometric calibration, and source extraction\footnote{\href{https://lco.global/documentation/data/BANZAIpipeline/}{https://lco.global/documentation/data/BANZAIpipeline/}}. We scheduled most observations in red bandpasses (I, i', z) where the S/N is highest for M dwarfs. Observing windows were chosen to include the full transit along with 1-3 hours of pre- and post-transit baseline. Many of our observations were defocused, to allow longer integration times for brighter stars and to smear the PSF over more pixels, reducing any error introduced by uncertainties in the flat-field. 

We performed differential aperture photometry on the data using the AstroImageJ tool \citep{Collins2017}. Using a finder chart, we drew apertures of varying radii (see Table \ref{tab:obs}) around the target star, 2-6 bright comparison stars, and any stars of similar brightness within 2.5'.~Light curves of the nearby stars were examined for evidence of being eclipsing binaries, variable stars, or the true source of the transit signal in \textit{TESS}' large pixels. For both of these systems, the transit was found around the target star, and no evidence of nearby eclipsing binaries or periodic stellar variation was found within 2.5' that could have given rise to the transit signal.

\textbf{TRAPPIST-South Photometry}

TRAPPIST-South at ESO-La Silla Observatory in Chile is a 60~cm Ritchey-Chretien telescope, which has a thermoelectrically cooled 2k$\times$2k FLI Proline CCD camera with a field of view of $22\arcmin\times22\arcmin$ and pixel-scale of 0.65\arcsec/px \citep{jehin2011,gillon2013}. We carried out a full-transit observation of TOI 122 on 2019 November 02 with $I+z$ filter with an exposure time of 60~s. We took 222 images and made use of AstroImageJ to perform aperture photometry, using an aperture radius of 8~pixels (5.2\arcsec) given the target PSF of 3.7\arcsec. We confirmed the event on the target star on time and we cleared all the stars of eclipsing binaries within the 2.5\arcmin~around the target star. For TOI 237 the observations were carried out on 2019 June 02 with $I+z$ filter and exposure time of 60~s. We took 207 images and used AstroImageJ to perform the aperture photometry, using an aperture radius of 8~pixels (5.2\arcsec) given the target PSF of 4.3\arcsec.

\subsection{SOAR Speckle Imaging}

High-angular resolution imaging is needed to search for nearby sources not resolved in the seeing-limited ground-based photometry. Nearby sources can contaminate the TESS photometry, resulting in a diluted transit and an underestimated planetary radius. We searched for nearby sources to TOI 122 with SOAR speckle imaging \citep{Tokovinin2018} on 2018 December 21 in I-band, a similar visible bandpass as TESS. Further details of observations from the SOAR TESS survey are available in \citet{Ziegler2020}.  We detected no nearby stars within 3\arcsec~of TOI 122 within the 5$\sigma$ detection sensitivity of the observation, which is plotted along with the speckle auto-correlation function in Figure \ref{fig:speckle}. Companions within 2.5 magnitudes of the target (which could dilute transit depths by 10\%) are excluded down to separations of about 0.3".

\subsection{Stellar Spectra}

\textbf{Magellan Spectra}

We obtained near-IR spectra of TOI 122 and TOI 237 on 2018 December 22 with the Folded-port InfraRed Echellete (FIRE) spectrograph \citep{simcoe2008}. FIRE is hosted on the 6.5 Baade Magellan telescope at Las Campanas Observatory. It covers the 0.8-2.5 micron band with a spectral resolving power of R = 6000. Both targets were observed in the ABBA nod patterns using the 0.6\arcsec~slit. TOI 122 was observed three times and TOI 237 was observed twice, both at 160s integration time. A nearby A0V standard was taken for both targets in order to aid with telluric corrections. The reduction of the spectra were completed using the FIREhose IDL package\footnote{\href{http://web.mit.edu/rsimcoe/www/FIRE/}{http://web.mit.edu/rsimcoe/www/FIRE/}}.

\textbf{SALT--HRS Spectra}

We obtained optical echelle spectra for each system using the High-Resolution Spectrograph \citep[HRS;][]{Crauseetal2014} on the Southern African Large Telescope \citep[SALT;][]{Buckleyetal2006}. Two observations were made for each system (TOI 122 on 2019 August 09, 10; TOI 237 on 2019 August 10, 12), with each epoch consisting of 3 consecutive integrations in the high-resolution mode ($R\sim$ 46,000). The spectra were reduced using a HRS-tailored reduction pipeline \citep{KniazevMN482016, KniazevSALT2017}\footnote{\href{http://www.saao.ac.za/$\sim$akniazev/pub/HRS\_MIDAS/HRS\_pipeline.pdf}{http://www.saao.ac.za/$\sim$akniazev/pub/HRS\_MIDAS/}}, which performed flat fielding and wavelength calibration. Due to the faint apparent magnitudes of these systems, we focused our analysis on wavelengths greater than 5000 \AA, where the spectra had signal-to-noise $>$ 10.

To determine systemic radial velocities for both systems and to search for spatially-unresolved stellar companions, we computed spectral-line broadening functions (BFs) for each observation. The BF is computed via a linear inversion of the observed spectrum with a narrow-lined template, and represents a reconstruction of the average photospheric absorption-line profile \citep{Rucinski1992,Tofflemireetal2019}. For both systems, the BF is very clearly single peaked, indicating a contribution from only one star. Figure \ref{fig:bf} presents a region of the SALT--HRS spectrum for each system with its corresponding template and broadening function.

For each spectrum, the BFs computed for each echelle order were combined and fit with a Gaussian profile to determine the system's radial velocity. Uncertainties on these measurements were derived from the standard deviation of the line fits for BFs combined from three independent subsets of the echelle orders. The radial velocity for each epoch was then calculated as the error-weighted mean of the three consecutive measurements from each night. More detail on this process can be found in \citet{Tofflemireetal2019}. From the two epochs spaced one to two days apart, we found no evidence for radial-velocity variability. The mean and standard error of the RV measurements are provided in Tables \ref{tab:sys_params1} and \ref{tab:sys_params2}.

\section{False Positive Vetting}

\textbf{Instrumental effects or statistical false positive}

From the SPOC data validation reports, the \textit{TESS} detections are significant with a S/N of 8.0 for TOI 122b and 9.8 for TOI 237b. These are both near the 7-$\sigma$ detection significance cutoff \citep{Jenkins2002}, which means these planets were found near \textit{TESS}' observational limits of discovery. However, given that we redetected transits of both planets from the ground, with consistent depths and timing,  we are confident these detections are in fact robust.

\begin{figure}[t!]
\centering
\includegraphics[width=0.47\textwidth]{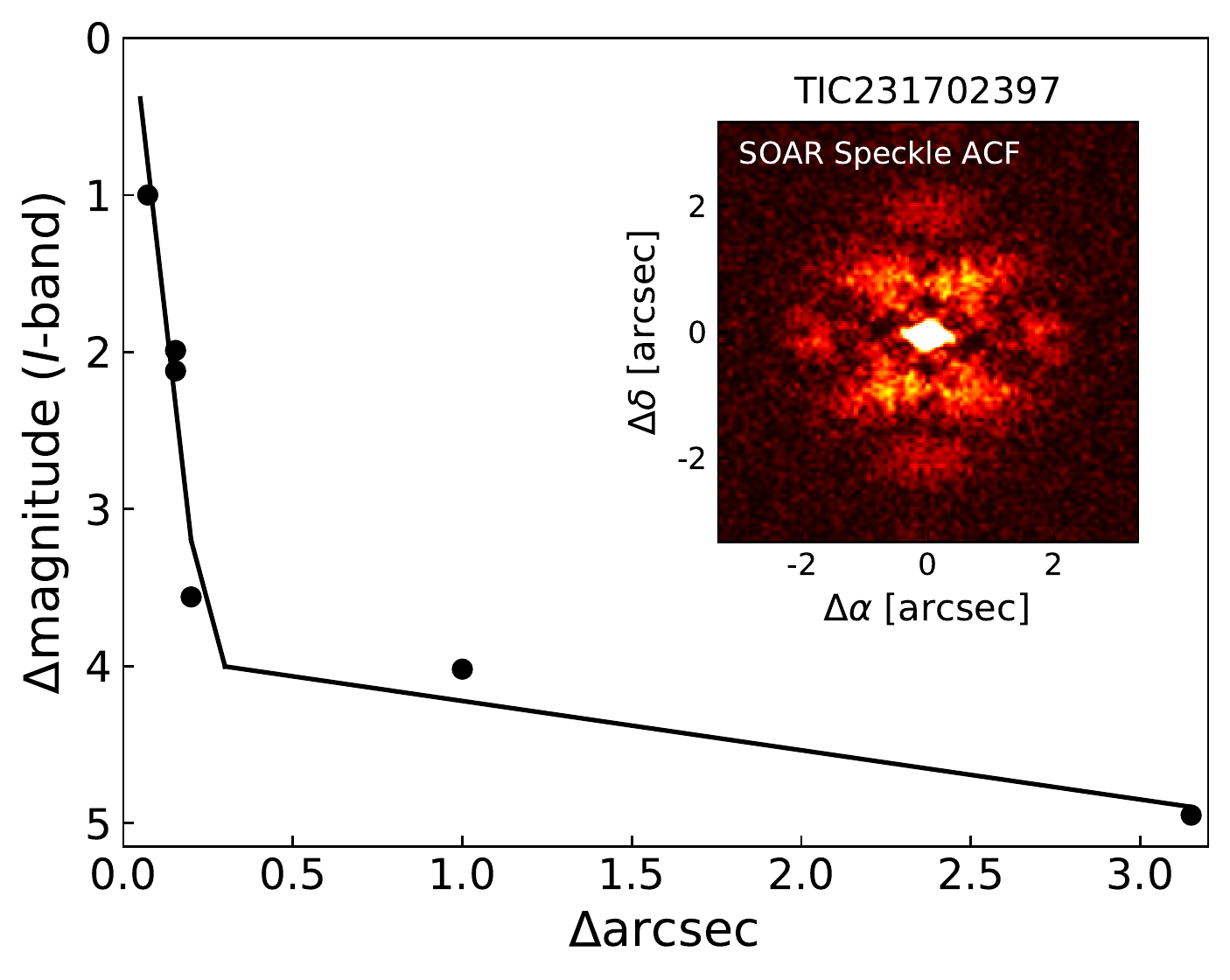}
\caption{5$\sigma$ detection limits of SOAR Speckle imaging for TOI 122. The inset shows that no companions were detected down to a limit of 3''.}
\label{fig:speckle}
\end{figure}

\textbf{Nearby transit or eclipsing binary}

For both of these planets, we searched all nearby ($<$ 2.5\arcmin~radius) stars in the seeing-limited LCO data that were bright enough to have caused the detected transits if blended in the \textit{TESS} photometry. We found no evidence of sources that were variable or eclipsing on the time scale of these planets' orbital periods. Both of these stars have high proper motions, and examination of archival images indicated that there are no bright stars at the targets' locations (see Fig. \ref{fig:finder}). In addition, we positively detected a transit in the aperture placed around the target star, so we believe these detections are not due to any physically-unbound nearby stars.

\begin{figure*}[t!]
\centering
\includegraphics[width=0.9\textwidth]{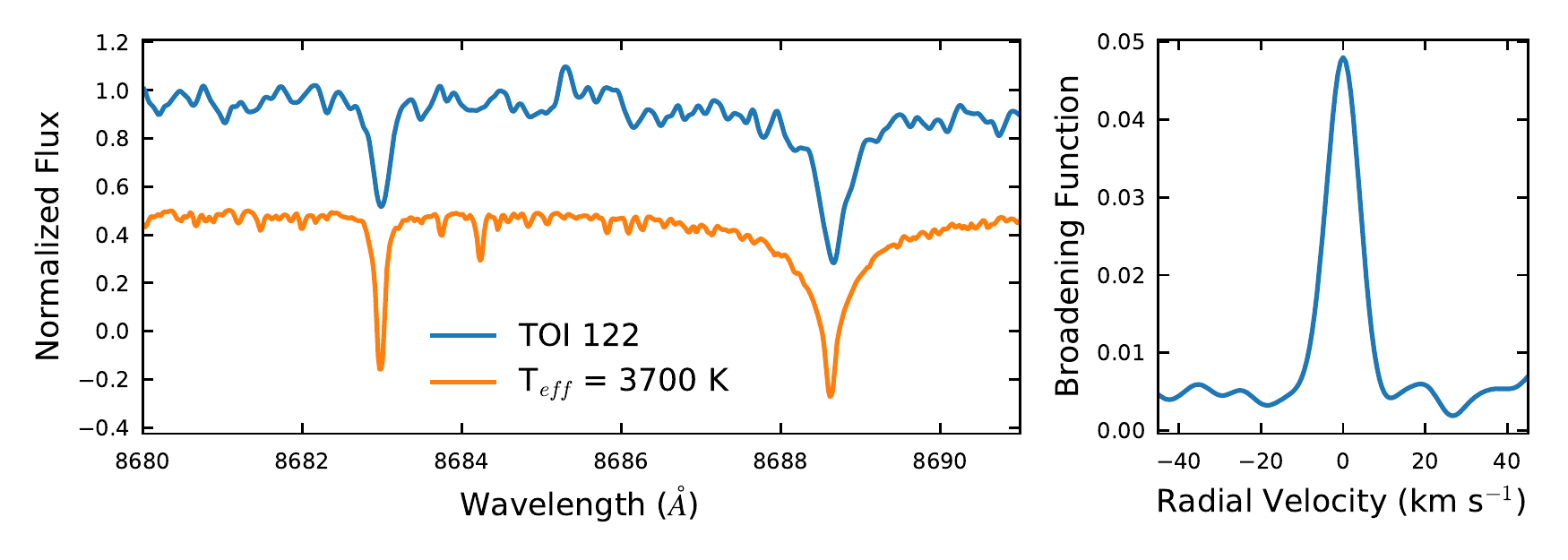}
\includegraphics[width=0.9\textwidth]{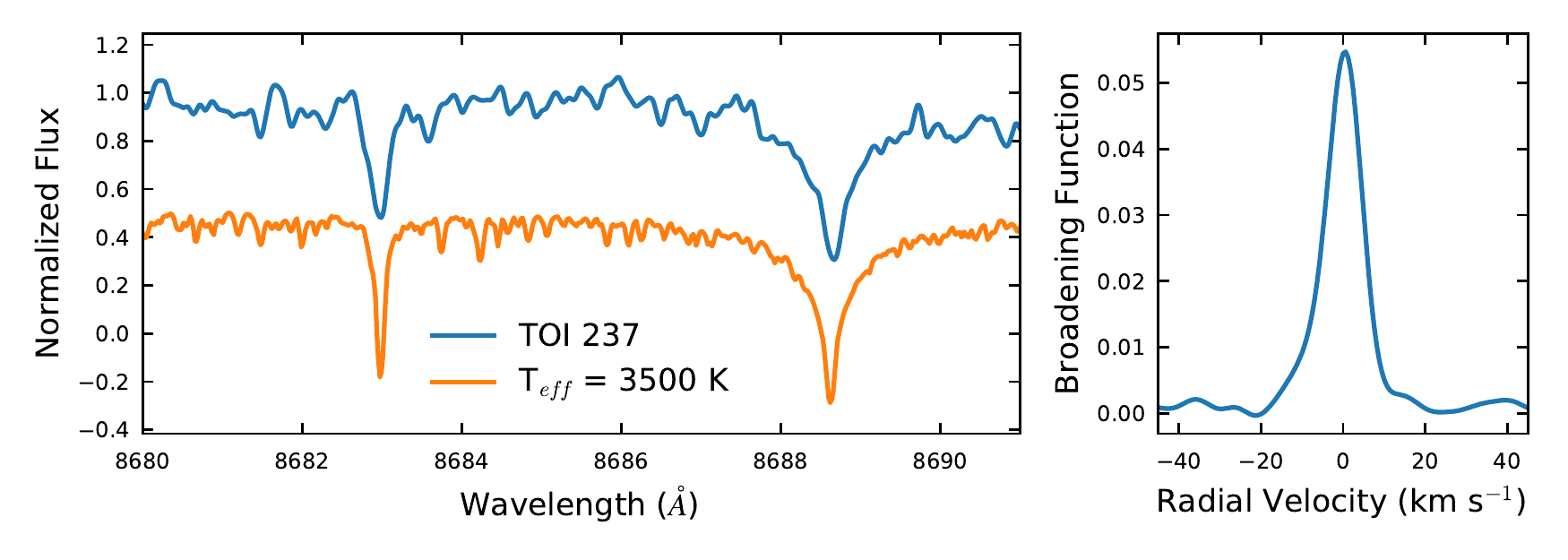}
\caption{{\bf Left:} Region of a SALT--HRS spectrum ({\em blue}) with the corresponding synthetic template ({\em orange}), where we have offset the flux slightly for clarity. {\bf Right:} The broadening function computed from this spectral region. Inspection of the broadening function and individual spectral lines indicates each system is single-lined, and does not host a short-period stellar companion. Note that the model temperatures cited on the figure are higher than the values we report for these two stars; this is discussed in Section \ref{section:stellar-params}}.
\label{fig:bf}
\end{figure*}

\textbf{Contaminated apertures}

The photometric apertures we used for the ground-based observations were typically $<$6\arcsec~(see Table \ref{tab:obs}), so we can rule out contaminating sources outside that approximate radius from our target stars. In the \textit{TESS} data, the PDCSAP light curves have already been corrected for contamination of nearby sources present in the TIC, and our higher-resolution ground-based observations show depths consistent with the \textit{TESS} light curves. SALT spectra show both sources to be single-lined, indicating a lack of evidence for unresolved luminous companions (see Fig. \ref{fig:bf}). We also obtained SOAR speckle imaging of TOI 122 which indicated there was not a nearby companion down to a separation of 0.3\arcsec~which could contaminate the aperture (see Fig. \ref{fig:speckle}).

\begin{figure*}[htbp]
\centering
\includegraphics[width=\textwidth]{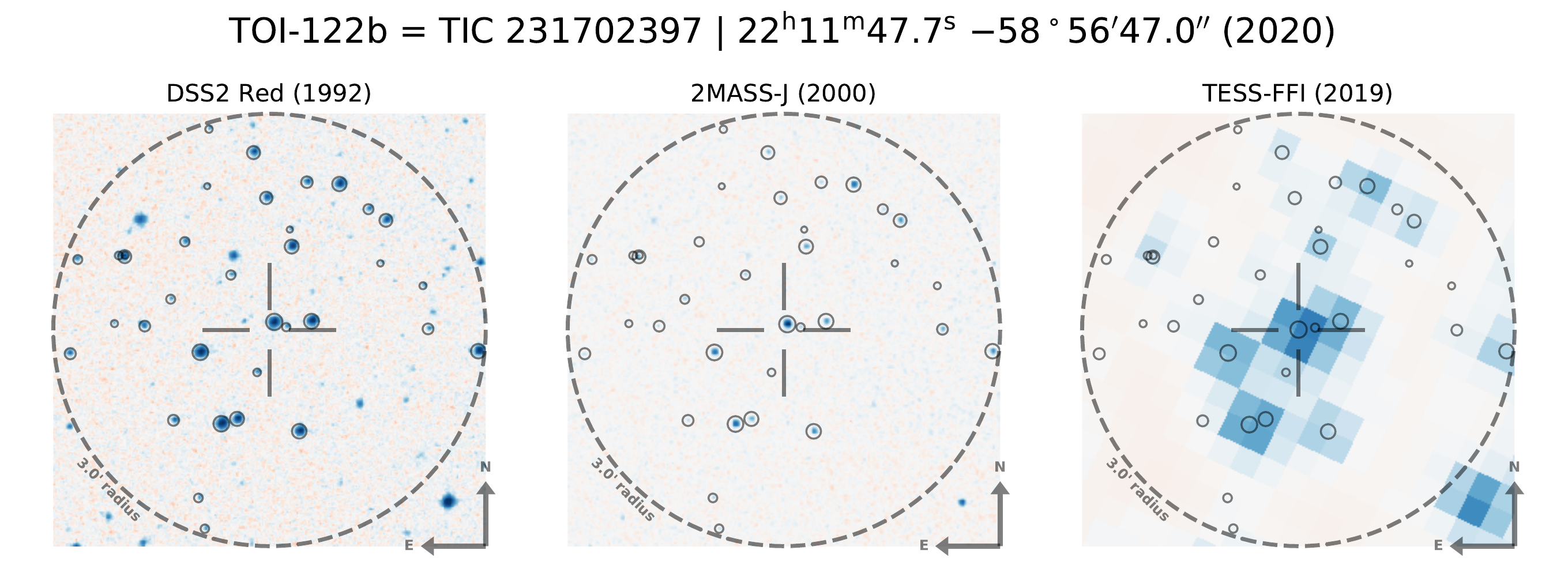}

\vspace{1cm}

\includegraphics[width=\textwidth]{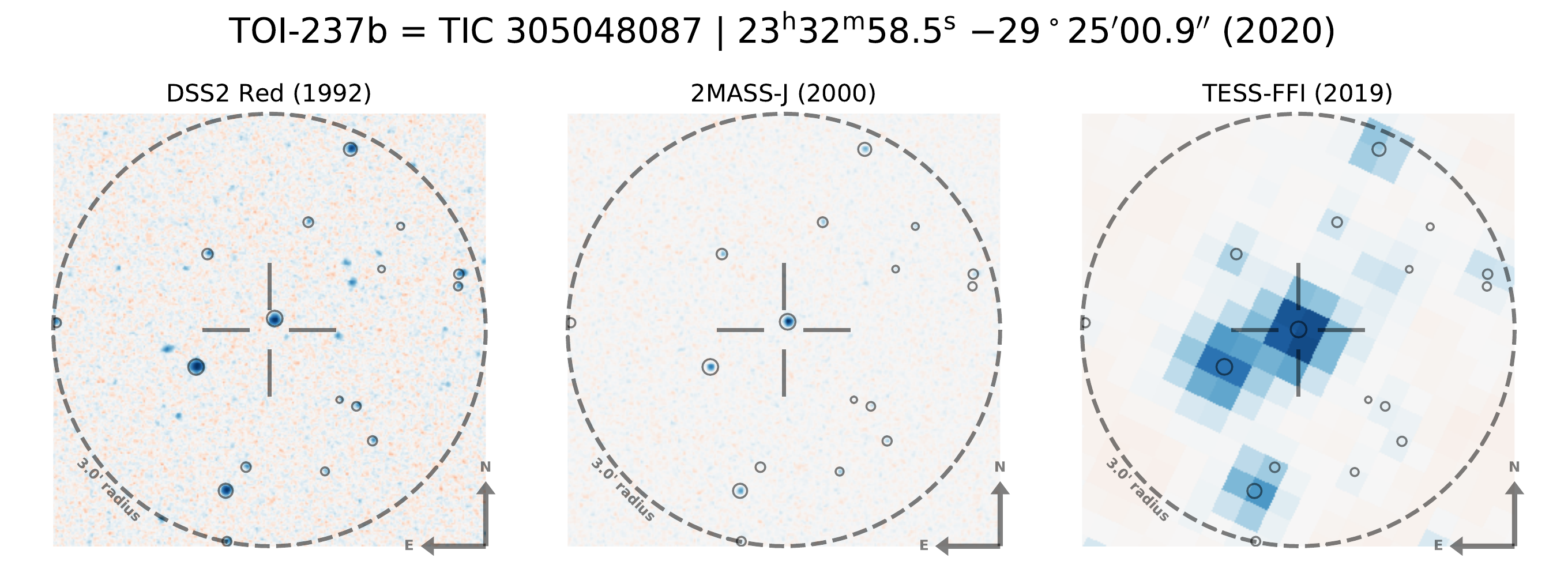}
\caption{Finder charts for TOI 122 ({\em top}) and TOI 237 ({\em bottom}), including scanned red-sensitive photograph plates from the Digitized Sky Survey ({\em left}), 2MASS  ({\em middle}), and the \textit{TESS} full-frame images ({\em right}). Circles indicate stars from Gaia DR2, with areas logarithmically expressing apparent brightness. Crosshairs indicate targets' position in the year 2019, near the time of the \textit{TESS} imaging.}
\label{fig:finder}
\end{figure*}

\textbf{Non-planet transiting object}

\begin{figure*}[t!]
\centering
\includegraphics[width=0.49\textwidth]{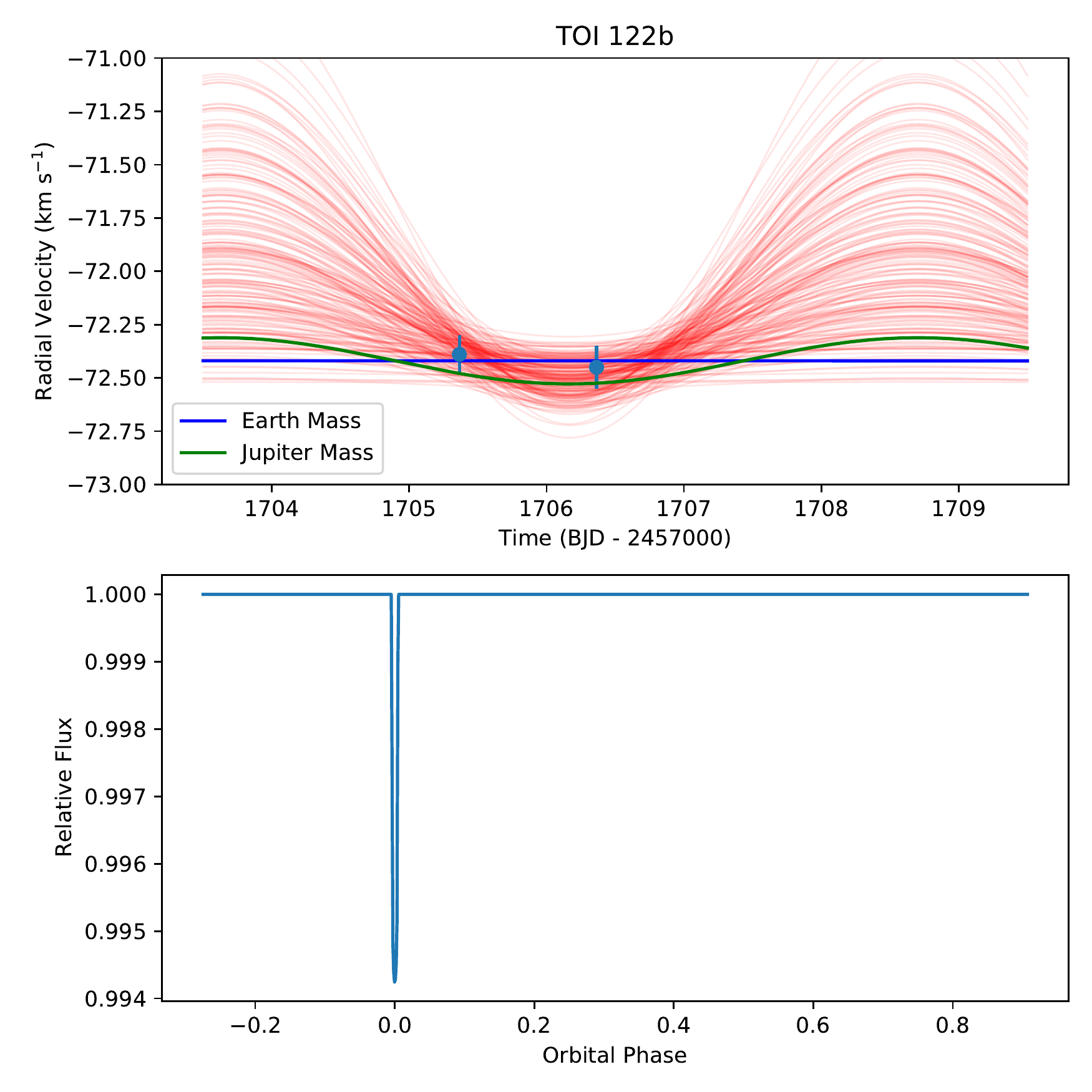}
\includegraphics[width=0.49\textwidth]{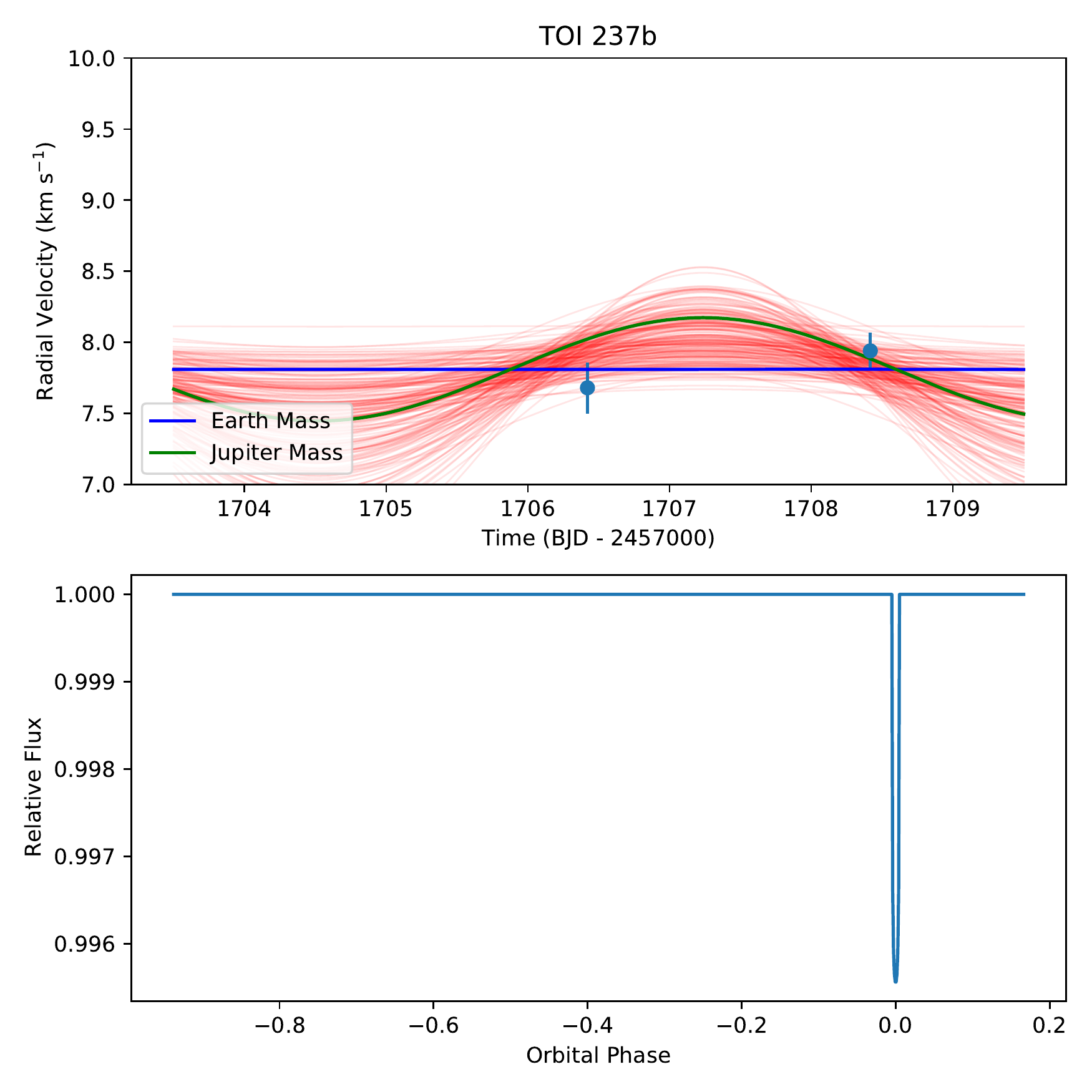}
\caption{Reconnaissance radial velocity observations from SALT--HRS for both systems, including model orbits for different planet masses ({\em top}) plotted with the corresponding transit light curves ({\em bottom}). Theoretical RV curves for Earth ({\em blue}) and Jupiter ({\em green}) masses are shown, as well as 200 random samples from the posterior distributions ({\em red}). While we cannot obtain precise planetary masses from these spectra, we are able to rule out super-planetary mass companions by calculating the maximum mass consistent with these measurements. These upper-limit masses based on the 95$^{th}$ percentile samples are 6.7 M$_{\rm{J}}$ for TOI 122b, and 2.1 M$_{\rm{J}}$ for TOI 237b.}
\label{fig:rvs}
\end{figure*}

Based on the measured transit depths and inferred stellar parameters, we can constrain both planets to R$_{\rm{p}}<$0.8~R$_{\rm{J}}$, which makes them small enough to be in the planet regime \citep{Burrows2011}. We also estimate upper-limit masses from the SALT radial velocity data. Using the two RV data points for each system, we model a range of masses consistent with these values to estimate the upper limit planet masses. These models were done using a 100k step MCMC (20k step burn-in) with the baseline and planet mass as free parameters, the assumption of circular orbits, and the only constraining prior that the planet mass is non-negative. We find the upper limit ($95^{th}$ percentile) masses for both of these planets to be in the planetary regime: M$_{\rm{p}}\leq6.7$~M$_{\rm{J}}$ for TOI 122b and $2.1$~M$_{\rm{J}}$ for TOI 237b (see Fig. \ref{fig:rvs}). Lastly, we have transit data in multiple bands for both objects, with consistent depths. This achromaticity suggests that these are non-luminous objects such as planets \citep[see][]{Parviainen2019}.

\section{Results}

\subsection{Light Curve Analysis}

For both systems, we omitted 2 observations of TOI 122 and 1 observation of TOI 237 where the transit is completely obscured by the noise. This corresponds to a photometric RMS such that the transit signal-to-noise is $\sim1$, which we argue is justified given the large number of observations which clearly show a transit (see Table~\ref{tab:obs}). We also omitted observations that did not capture the mid-transit, to prevent the MCMC walkers from running away with obviously incorrect mid-transit times and semi-major axes. We modeled all ground-based light curves simultaneously by requiring the inclination, a/R$_{\star}$, and R$_{\rm{p}}$/R$_{\star}$ to be the same value across all transits, but allowing T$_0$ to vary for transits at different epochs. T$_0$ is fixed between transits that occurred at the same epoch (where we have observations from multiple telescopes, for example). To fit the baseline flux alongside the light curve parameters, We implemented a linear 2-parameter airmass model of the form $(C_1+C_2a)B$ where $a$ is the airmass at each exposure and $B$ is the \texttt{BATMAN} light curve model. This added up to 24 modeled parameters for TOI 122b and 20 parameters for TOI 237b, the difference being due to a different number of observations for both systems. After analyzing the follow-up lightcurves and refining the orbital periods, we modeled the phase-folded \textit{TESS} light curves to examine how well the systems' properties were improved. For a discussion on period refinement, see \S \ref{section:Period Refinement}.

The models are created using \texttt{BATMAN} \citep{Kreidberg2015}, which is based on the analytic transit model from \citet{MandelandAgol}. Stellar limb darkening coefficients were calculated for each separate bandpass with \texttt{LDTk}, the stellar Limb Darkening Toolkit \citep{Parviainen2015}, and these coefficients are listed in Table \ref{tab:ldtk_params}. Figures \ref{TESSLightcurves}, \ref{TOI122_Lightcurves}, and \ref{TOI237_Lightcurves} show all transit light curves with models.

We found posterior distributions through Bayesian analysis using \texttt{emcee} \citep{Foreman-Mackey2012}. We ran the MCMC with 150 walkers and 200k steps, discarding the first 40k steps (20\%) and using uniform priors for all parameters. We chose the number of steps based on when each chain converged, using the integrated autocorrelation time heuristic built into \texttt{emcee}. With our 160k steps (post-burn-in), all chains reached $>$100 independent samples, suggesting adequate convergence \citep[for a discussion of MCMC convergence, see][]{Hogg2018}. The priors are set so that the planet does not have a negative radius ($0\leq$~R$_{\rm{p}}$/R$_\star\leq$~1), the mid-transit time is within the range of the data, the eccentricity is 0, the semi-major axis is physically reasonable ($2\leq$~a/R$_\star\leq$~200), and the inclination is geometrically limited to be i~$\leq$~90\degree to avoid duplicate solutions of i$>$90\degree.

\begin{figure*}[t!]
\centering
\subfloat[]{\includegraphics[width=0.59\textwidth]{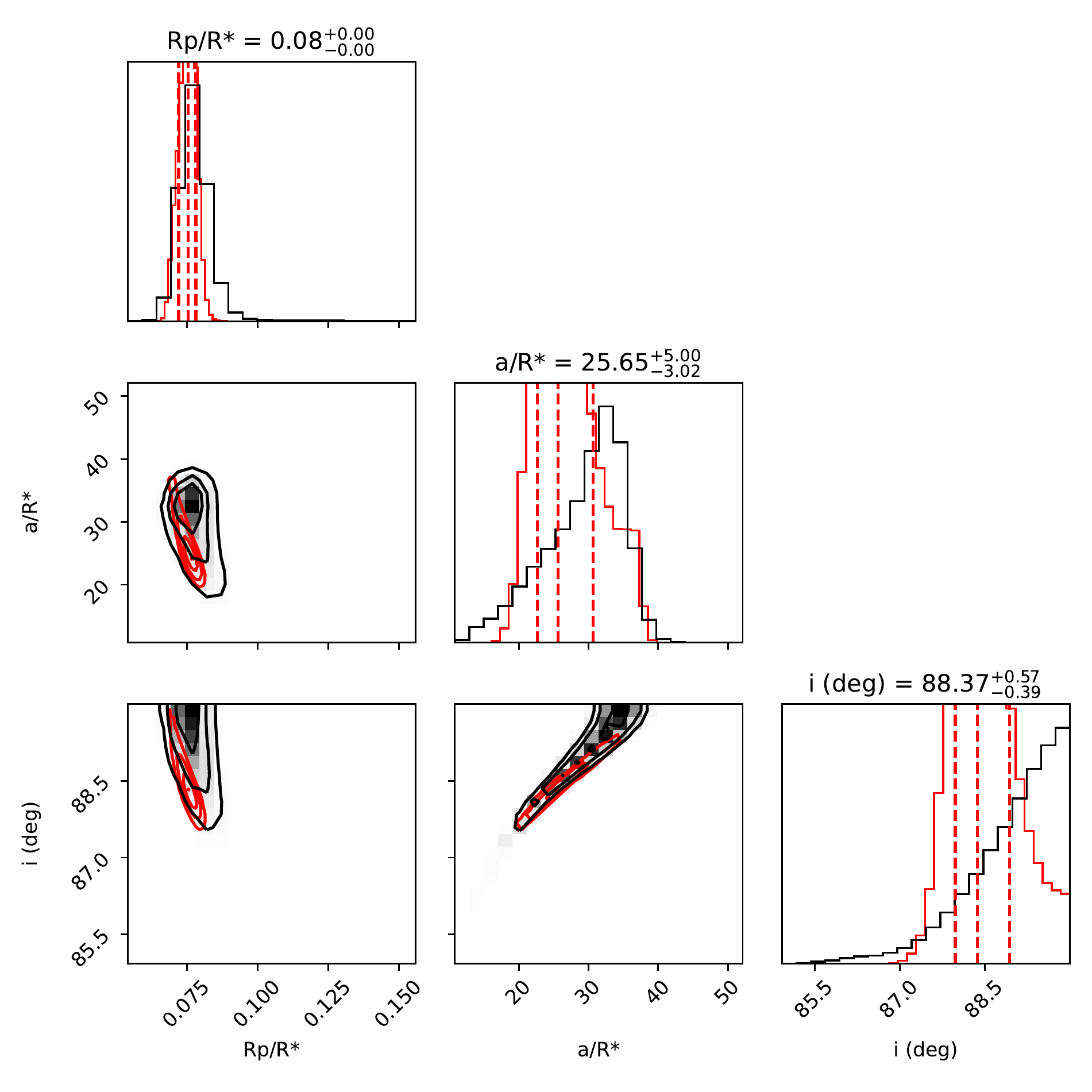}}\\
\subfloat[]{\includegraphics[width=0.59\textwidth]{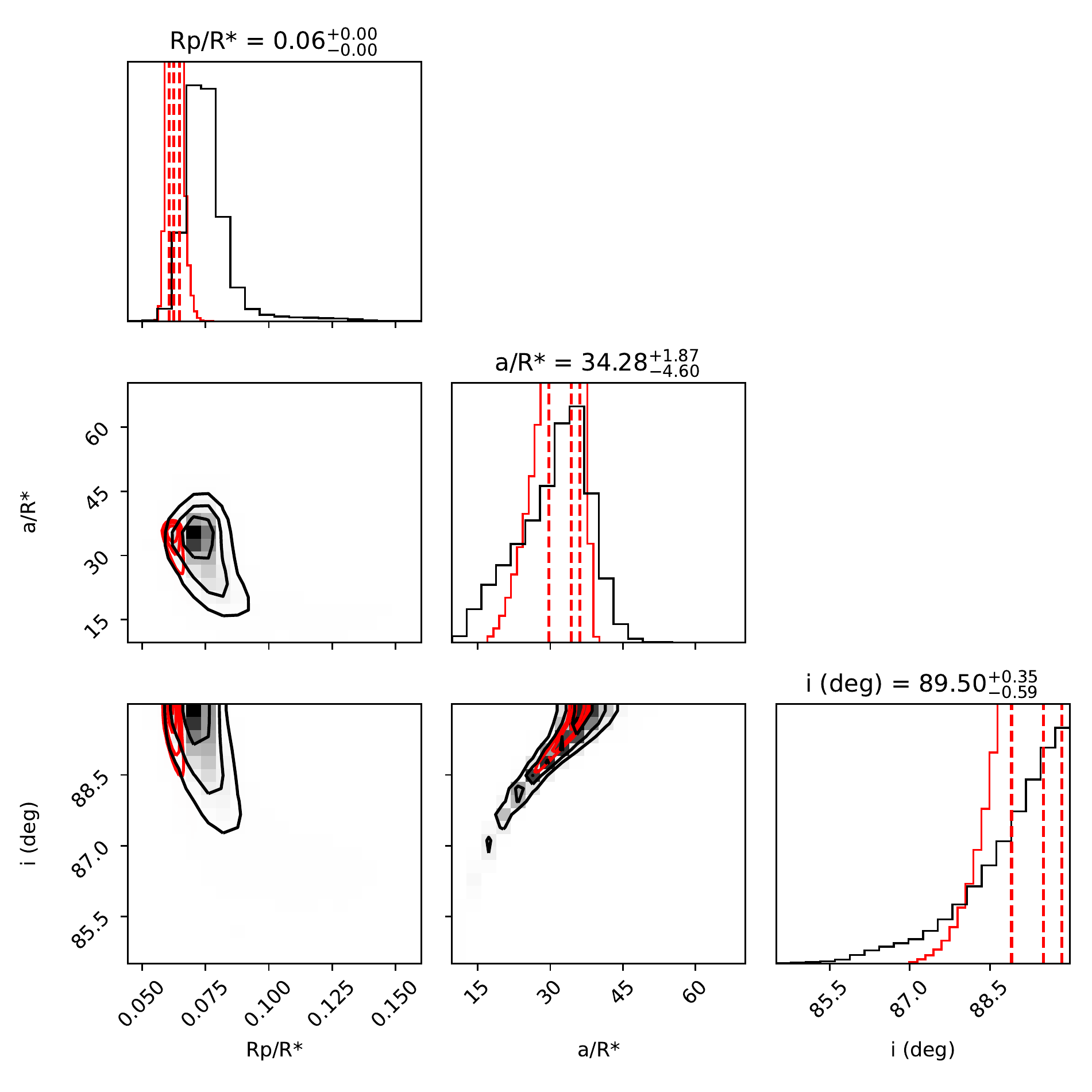}}
\caption{Corner plots \citep{corner} for the MCMC posteriors of all fits for TOI 122b (top) and TOI 237b (bottom). The posteriors from modeling only the phase-folded \textit{TESS} light curves ({\em gray}) agree with those from modeling only the ground-based follow-up light curves ({\em black}), with the constraints from ground-based telescopes being more precise due to their larger apertures. Labels on top of the posteriors are from the ground-based results. \label{Corner}}
\end{figure*}

The results cited in Tables \ref{tab:sys_params1} and \ref{tab:sys_params2} are the 50$^{th}$ percentile values with 1-$\sigma$ uncertainties based on the central 68\% confidence intervals of the ground-based MCMC samples which have had the burn-in removed. In Figure \ref{Corner}, we show the posterior distributions from fitting only the folded \textit{TESS} light curves as well as posterior distributions for only the follow-up transits, for both systems. Results from modeling the follow-up transits are consistent with the \textit{TESS} fits, but the ground-based follow-up provides much tighter constraints due to the improved signal-to-noise we get with the larger-aperture LCO 1-m telescopes and from having additional independent transits.

\subsection{Stellar Parameters}
\label{section:stellar-params}

\textbf{Mass and Radius:} We first used the emprical relations in \citet{Mann2019} to calculate stellar masses from Gaia parallaxes and 2MASS K-band magnitudes. From Gaia DR2 \citep{GaiaCollaboration2018}, the distance to TOI 122 is \TOIOneDist~pc and the distance to TOI 237 is \TOITwoDist~pc. Using the \citet{Mann2019} relations, we get M$_{\star}$~=~\TOIOneMs~M$_{\sun}$ for TOI 122 and M$_{\star}$~=~\TOITwoMs~M$_{\sun}$ for TOI 237. Using the analogous \citet{Mann2015} absolute M$_K$ relation for stellar radii, we found R$_{\star}$=\TOIOneRsMann~R$_{\sun}$~and~\TOITwoRsMann~R$_{\sun}$ for TOI 122 and 237, respectively. As a verification, we compared the stellar densities from the empirical masses and radii to the stellar densities calculated directly from the light curves:
\begin{equation}
    \rho_{\star} = \frac{3\pi}{GP^2}\left(\frac{a}{R_{\star}}\right)^3,
\end{equation}
where $\rho_{\star}$ is the stellar density, $P$ is the orbital period of the planet, $a/R_{\star}$ is the normalized semi-major axis, and we have assumed circular orbits \citep{Seager2003,Sozzetti2007}. The densities derived from the light curves are \TOIOneLCDens~g cm$^{-3}$ for TOI 122 and \TOITwoLCDens~g cm$^{-3}$ for TOI 237, which agree well with the densities from our empirically derived masses and radii (\TOIOneMannDens~g cm$^{-3}$ and \TOITwoMannDens~g cm$^{-3}$ for TOI 122 and 237, respectively). Similarly, we calculated the semi-major axes of these systems from the stellar mass predictions and measured periods, and convert them to a/R$_\star$ using the \citet{Mann2015} empirically predicted radii. These calculated semi-major axes give us a/R$_{\star}$ of \TOIOneaRs~(compared to \TOIOneaRsLC~from the light curves) and \TOITwoaRs~(compared to \TOITwoaRsLC~from the light curves) for TOI 122b and 237b.

\textbf{Effective Temperature (\teff) and Luminosity:} For both stars,we calculated \teff~using six of the different empirical color magnitude relations (equations 1-3 and 11-13 of Table 2) in \citet{Mann2015}. Taking the weighted average of the six temperatures, we get \teff~= \TOIOneTeff~K for TOI 122 and \TOITwoTeff~K for TOI 237. For both sets of calculations, the standard deviation of the six temperatures was $\sim55$~K.

For stellar luminosities, we calculate the V-band bolometric correction based on the V-J empirical relation in \citet{Mann2015}. This gives luminosities of \TOIOneLsun~L$_{\sun}$ and \TOITwoLsun~L$_{\sun}$ for TOI 122 and 237, respectively. We then compared these luminosities to the luminosities calculated from the \citet{Mann2015} radii and effective temperatures (described above):
\begin{equation}
    \frac{L}{L_\sun} = \left(\frac{R}{R_\sun}\right)^2\left(\frac{T_{\rm{eff}}}{T_\sun}\right)^4,
\end{equation}
where we use T$_{\sun}=5772$~K \citep{SolarValues}. This resulted in L$=$0.013$\pm$0.003~L$_{\sun}$ for TOI 122 and L$=$0.0042$\pm$0.0007~L$_{\sun}$ for TOI 237, in good agreement with the bolometric-correction luminosities. Given the collective agreement between light curve densities, bolometric luminosities, and empirical estimates for radii, masses, and effective temperatures, we adopt the \citet{Mann2015,Mann2019}-derived stellar parameters and corresponding uncertainties for these two stars.

\begin{table*}
\begin{center}
\begin{tabular}{ccccc}
\toprule
    Parameter & Value & Source \\
    \midrule
    \textbf{TOI 122} \\
    TIC ID & 231702397 & TICv8\\
    RA (J2000) & \TOIOneRa & TICv8\\
    Dec (J2000) & \TOIOneDec & TICv8\\
    TESS Magnitude & $13.048\pm0.007$ & TICv8 \\
    Apparent V Magnitude & $15.526\pm0.026$ & TICv8 \\
    Apparent J Magnitude & $11.531\pm0.024$ & TICv8 \\
    Apparent H Magnitude & $11.020\pm0.022$ & TICv8 \\
    Apparent K Magnitude & $10.771\pm0.021$ & TICv8 \\
    Gaia DR2 ID & 6411096106487783296 & Gaia DR2 \\
    Distance [pc] & \TOIOneDist & Gaia DR2 \\
    Proper Motion RA [mas yr$^{-1}$] & \TOIOnePMRA & Gaia DR2 \\
    Proper Motion DEC [mas yr$^{-1}$] & \TOIOnePMDEC & Gaia DR2 \\
    Gaia G mag & 14.3357 & Gaia DR2 \\
    Gaia RP mag & 13.1523 & Gaia DR2 \\
    Gaia BP mag & 15.7971 & Gaia DR2 \\
    Stellar Mass [M$_\Sun$] & \TOIOneMs & Derived from \citet{Mann2019} \\
    Stellar Radius [R$_\Sun$] & \TOIOneRsMann & Derived from \citet{Mann2015} \\
    T$_{\rm{eff}}$ [K] & \TOIOneTeff & Derived from \citet{Mann2015}\\
    Luminosity [L$_{\odot}$] & \TOIOneLsun & Derived from \citet{Mann2015} \\
    Stellar log $g$ & $4.88\pm0.05$ & This Work \\
    Radial Velocity [km~s$^{-1}$] & \TOIOneRv & This Work\\
    Stellar Density [g~cm$^{-3}$] & \TOIOneLCDens & This Work \\
    v~sin\textit{i} [km s$^{-1}$] & $\leq7.2$ & This Work \\
    H$\alpha$ Equivalent Width [\rm{\AA}] & $0.09$ & This Work \\
    \midrule
    \textbf{TOI 122b} \\
    Period [days] & \TOIOnePeriod & This Work \\
    Transit Depth [\%] & 0.56 & This Work \\
    R$_{\rm{p}}$/R$_{\star}$ & \TOIOneRpRs & This Work \\
    Planet Radius [R$_\Earth$] & \TOIOneRp & This Work \\
    Planet Mass [M$_\Earth$] & \TOIOneMp & Predicted from \citet{Chen2017} \\
    Planet Type & 100\% Neptunian & Predicted from \citet{Chen2017} \\
    $\frac{a}{R_\star}$ & \TOIOneaRs & This Work\\
    Semi-major Axis [AU] & $0.0392\pm0.0007$ &  This Work \\
    \textit{i} [degrees] & \TOIOnei & This Work\\
    Impact Parameter (b) & \TOIOneb & This Work\\
    Insolation [S$_\Earth$] & \TOIOneIns & This Work\\
    Equilibrium Temperature, T$_{\rm{eq}}$ [K]: & & \multirow{ 4}{*}{This Work} \\
    Bond Albedo = 0.75 (Venus-like) & \TOIOneTeqLow & \\
    Bond Albedo = 0.3 (Earth-like) & \TOIOneTeqHigh & \\
    Bond Albedo = 0 (Upper Limit) & \TOIOneTeqUpper & \\
\bottomrule
\end{tabular}
\end{center}
\caption{System parameters for TOI 122b. TICv8 information can be found in \citet{TICv8}.}
\label{tab:sys_params1}
\end{table*}

\begin{table*}
\begin{center}
\begin{tabular}{ccccc}
\toprule
    Parameter & Value & Source \\
    \midrule
    \textbf{TOI 237} \\
    TIC ID & 305048087 & TICv8\\
    RA (J2000) & \TOITwoRa & TICv8\\
    Dec (J2000) & \TOITwoDec & TICv8\\
    TESS Magnitude & $13.410\pm0.007$ & TICv8 \\
    Apparent V Magnitude & $16.37\pm0.20$ & TICv8 \\
    Apparent J Magnitude & $11.74\pm0.02$ & TICv8 \\
    Apparent H Magnitude & $11.019\pm0.022$ & TICv8 \\
    Apparent K Magnitude & $10.896\pm0.025$ & TICv8 \\
    Gaia DR2 ID & 2329387852426700800 & Gaia DR2 \\
    Distance [pc] & \TOITwoDist & Gaia DR2 \\
    Proper Motion RA [mas yr$^{-1}$] & \TOITwoPMRA & Gaia DR2 \\
    Proper Motion DEC [mas yr$^{-1}$] & \TOITwoPMDEC & Gaia DR2 \\
    Gaia G mag & 14.754 & Gaia DR2 \\
    Gaia RP mag & 13.5016 & Gaia DR2 \\
    Gaia BP mag & 16.4447 & Gaia DR2 \\ 
    Stellar Mass [M$_\Sun$] & \TOITwoMs & Derived from \citet{Mann2019} \\
    Stellar Radius [R$_\Sun$] & \TOITwoRsMann &  Derived from \citet{Mann2015} \\
    T$_{\rm{eff}}$ [K] & \TOITwoTeff & Derived from \citet{Mann2015}\\
    Luminosity [L$_{\odot}$] & \TOITwoLsun & Derived from \citet{Mann2015} \\
    Stellar log $g$ [cgs] & $5.04\pm0.07$ & This Work \\
    Radial Velocity [km~s$^{-1}$] & \TOITwoRv & This Work\\
    Stellar Density [g~cm$^{-3}$] & \TOITwoLCDens & This Work \\
    v~sin\textit{i} [km s$^{-1}$] & $\leq6.4$ & This Work \\
    H$\alpha$ Equivalent Width [\rm{\AA}] & $1.74$ & This Work \\
    \midrule
    \textbf{TOI 237b} \\
    Period [days] & \TOITwoPeriod & This Work \\
    Transit Depth [\%] & 0.38 & This Work \\
    R$_{\rm{p}}$/R$_{\star}$ & \TOITwoRpRs & This Work \\
    Planet Radius [R$_\Earth$] & \TOITwoRp & This Work \\
    Planet Mass [M$_\Earth$] & \TOITwoMp & Predicted from \citet{Chen2017} \\
    Planet Type & 25\% Terran, 75\% Neptunian & Predicted from \citet{Chen2017} \\
    $\frac{a}{R_\star}$ & \TOITwoaRs & This Work\\
    Semi-major Axis [AU] & $0.0341\pm0.0010$ & This Work \\
    \textit{i} [degrees] & \TOITwoi & This Work\\
    Impact Parameter (b) & \TOITwob & This Work\\
    Insolation [S$_\Earth$] & \TOITwoIns & This Work\\
    Equilibrium Temperature, T$_{\rm{eq}}$ [K]: & & \multirow{ 4}{*}{This Work} \\
    Bond Albedo = 0.75 (Venus-like) & \TOITwoTeqLow & \\
    Bond Albedo = 0.3 (Earth-like) & \TOITwoTeqHigh & \\
    Bond Albedo = 0 (Upper Limit) & \TOITwoTeqUpper & \\
\bottomrule
\end{tabular}
\end{center}
\caption{System parameters for TOI 237b.}
\label{tab:sys_params2}
\end{table*}

\begin{table}
\begin{center}
\begin{tabular}{ccc}
\toprule
      Filter & Value [$u_1,u_2$] & Uncertainty [$\sigma_1,\sigma_2$]\\
    \midrule
    \multicolumn{2}{c}{\textbf{TOI 122}} \\
    V & [0.5266, 0.2934]&[0.0151, 0.0240]\\
    g' & [0.5161, 0.2998]&[0.0124, 0.0200]\\
    r' & [0.5209, 0.2644]&[0.0149, 0.0234]\\
    i' & [0.3050, 0.2898]&[0.0069, 0.0139]\\
    I & [0.2558, 0.2566]&[0.0046, 0.0098]\\
    I\&z' & [0.2768, 0.2918]&[0.0067, 0.0140] \\
    \midrule
    \multicolumn{2}{c}{\textbf{TOI 237}} \\
     g' & [0.5720, 0.2925]&[0.0191, 0.0296] \\
     I & [0.2657, 0.2911]&[0.0100, 0.0205]\\
     I\&z' & [0.2967, 0.3343]&[0.0138, 0.0260]\\
\bottomrule
\end{tabular}
\end{center}
\caption{Quadratic limb darkening parameters [$u_1,u_2$] and associated uncertainties [$\sigma_1,\sigma_2$], calculated using \texttt{LDTk} using the stellar parameters listed in Tables \ref{tab:sys_params1} and \ref{tab:sys_params2}.}
\label{tab:ldtk_params}
\end{table}

We chose to calculate our stellar parameters based on empirical models rather than adopting values from our spectral observations because of some inconsistencies in the spectra. The method we used to analyze RV signals from SALT spectra is optimized to detect precise RVs but not to accurately calculate stellar temperature. Therefore, the temperature that corresponds to the best fit RV model is not necessarily an accurate estimate of stellar temperature. This aspect of the modeling does not affect the $vsini$ values presented in this paper. The FIRE spectra indicate TOI 122 is a significantly larger and hotter M dwarf, opposing other estimates of its size and temperature. We attribute this to the observing conditions and telluric contamination of the Magellan FIRE spectra, and we therefore do not use the effective temperatures and radii we derive from these spectra.

\subsection{Assumption of Circular Orbits}

All of the analysis was done under the assumption of circular orbits for these two systems. To justify this, we calculate the tidal circularization timescales following \citet{Goldreich1966}:
\begin{equation}
    \tau_{circ}=\frac{2PQ'}{63\pi}\left(\frac{M_p}{M_\star}\right)\left(\frac{a}{R_p}\right)^5,
\end{equation}
where $P$ is the planet's orbital period and $Q'$ quantifies how well the planet dissipates energy under deformation. Rocky planets tend to have lower $Q'$ values while gaseous planets have larger $Q'$ values. We adopt $Q'=1\times10^4$ for TOI 122b and $Q'=500$ for TOI 237b. These values are based on $Q'$ values derived for the solar system planets, where Earth has $Q'\sim$100 and Neptune has a $Q'\sim6\times10^4$ \citep{Goldreich1966}. We do not have measurements of $M_p$ for these planets, but our predicted masses based on the empirical relations in \citet{Chen2017} provide a precise enough estimate for this timescale. For TOI 122b and 237b, we calculate $\tau_{\rm{circ}}$ of 0.59 Gyr and 0.17 Gyr, respectively.

From the SALT spectra, we derived upper limits on $v$sin$i$ to be $<$7.2 km~s$^{-1}$ for TOI 122 and $<$6.4 km~s$^{-1}$ for TOI 237, which allow us to derive lower limits on the rotational periods of both stars under the assumption that the stellar rotation axis is perpendicular to the line of sight. We find those lower limits to be $>2.3$ days for TOI 122 and $>1.7$ days for TOI 237. In addition, the lack of any significant flaring activity or rotational modulation seen in the \textit{TESS} light curves for these two systems leads us to assume the stellar rotational periods are long, and probably greater than 27 days (the \textit{TESS} observation window for a single sector). While the relation between rotation period and age for M dwarfs is poorly constrained, \citet{Newton2016b} found the rotation rates of field M dwarfs to be between 0.1 and 140 days, with M dwarfs younger than 2 Gyr having rotational periods less than 10 days. We also calculate the H$\alpha$ equivalent widths (EW) from the SALT spectra, as H$\alpha$ emission is indicative of the activity level of M dwarfs \citep[see][]{Newton2017}. We find the EWs to be 0.09 \rm{\AA} for TOI 122 and 1.74 \rm{\AA} for TOI 237, placing both of these stars in the canonically inactive regime (EW$>$-1\rm{\AA}). \citet{Newton2017} provide a more direct way to estimate the rotational periods of inactive M dwarfs based on a polynomial fit with stellar mass. Given our derived masses for these two stars, we predict P$_{122}=72\pm22$~d and P$_{237}=102\pm22$~d from that relation. From the age-inactivity-spectral type relationship for cool stars described in \citet{West2008}, we predict that TOI 122 (an M3V) is likely older than 2 Gyr, and TOI 237 (an M4.5V) \citep[spectral types based on][]{Rajpurohit2013} is likely older than 4.5 Gyr, consistent with our other estimates of their ages.

We can see a picture emerging that these stars are inactive, slowly rotating, and old, in spite of precise stellar ages being difficult to obtain for M dwarfs. Given that $\tau_{\rm{circ}}$ for both planets is $<1$~Gyr, we assume both planets are on circular orbits. Our assumption that eccentricity is $\sim$ 0 is also supported by the agreement between the stellar densities calculated from the light curves and densities based on empirical estimates of mass and radius (see Section \ref{section:stellar-params}).

\subsection{Insolation and T$_{eq}$}

In order to form a picture of the thermal environment of these planets, we calculate the insolation these planets receive, relative to the bolometric flux that Earth receives from the Sun. We also calculate equilibrium temperatures under different assumptions for the Bond albedo, $A_{\rm B}$, which is the fraction of incident stellar radiation that is reflected by the planet, integrated over both wavelength and angle.

Under the assumptions of circular orbits, efficient heat redistribution, and planets that are thermal emitters \citep[for a discussion of these assumptions, see][]{Cowan2011}, we use the a/R$_{\star}$ values derived from our orbital periods and stellar masses to calculate planetary equilibrium temperature as:
\begin{equation}
    T_{\rm{eq}}=(1-A_{\rm B})^{\frac{1}{4}}\left(\frac{2a}{R_\star}\right)^{-\frac{1}{2}}T_{\rm{eff}},
\end{equation}
and insolation as:
\begin{equation}
    \frac{S}{S_{\Earth}} = \left(\frac{T_{\rm{eff}}}{T_\sun}\right)^4\left(\frac{a_\Earth/R_\sun}{a/R_\star}\right)^2,
\end{equation}
where $S$ is the bolometric insolation, $a$ is the semi-major axis derived from the stellar masses and orbital periods, $R_\star$ is the inferred stellar radius, and a$_\Earth$/R$_\sun=215$. We present $T_{\rm{eq}}$ (see Tables \ref{tab:sys_params1} and \ref{tab:sys_params2}) as a range of values assuming an Earth-like $A_{\rm B}=0.3$, a Venus-like $A_{\rm B}=0.75$, and $A_{\rm B}=0$.

\subsection{Period Refinement and TTVs}
\label{section:Period Refinement}

For both systems, we fit a linear model to the \textit{TESS} epoch and the follow-up epochs to refine the period, which we cite in Tables \ref{tab:sys_params1} and \ref{tab:sys_params2}. In doing this, we are also able to examine the difference between the expected and observed mid-transit times to search for evidence of periodic TTVs. The reduced-$\chi^2$ of a linear ephemeris ($2.2$ and $2.3$ for TOI 122b and 237b, respectively) gave marginal hints of variations on the time scale of minutes, but a Lomb-Scargle periodogram \citep[for a discussion of Lomb-Scargle periodograms, see][]{VanderPlas2018} applied to the O-C (observed minus calculated) mid-transit times showed no significant periodicity for either system, so we report no significant TTV detection.

\section{Discussion \& Conclusions}

These two planets help fill the parameter space for cool worlds near the boundary between rocky and gas-rich compositions. Neither is in the circumstellar habitable zone of its star as both receive more flux than the approximately 0.9 S$_{\Earth}$ moist greenhouse inner limit calculated by \citet{Kopparapu2013b} for stars with these effective temperatures. However, with insolations of \TOIOneIns~and \TOITwoIns~S$_{\Earth}$, they are relatively cool among known transiting exoplanets.

\subsection{Radial Velocity Prospects}

We do not have mass-constraining radial velocities for these two stars, so we applied the \citet{Chen2017} empirical mass-radius forecaster to predict M$_{\rm{122b}}=$\TOIOneMp~M$_{\Earth}$ and M$_{\rm{237b}}=$\TOITwoMp~M$_{\Earth}$, based on the planets' radii. The degeneracy between planet radius and bulk composition leads to large uncertainties in these predicted masses. The forecaster results classify TOI 122b as 100\% likely Neptunian and TOI 237b as 25\% likely to be Terran and 75\% likely to be Neptunian, where ``Terran" is the term used by \citet{Chen2017} to describe worlds similar to the inner terrestrial solar system planets and ``Neptunian" is used to describe worlds similar in their basic properties to Neptune and Uranus. The transition between these planet types was found by \citet{Chen2017} to be at 2.0$\pm$0.7 M$_{\earth}$. We can compare the stellar magnitudes and predicted RV semi-amplitudes to the current and near-future capabilities of RV facilities. Using the periods, stellar masses, and predicted planet masses, we estimate RV semi-amplitudes of 7.1 m s$^{-1}$ and 3.4 m s$^{-1}$ for TOI 122b and 237b, respectively. These semi-amplitudes are above the instrumental noise floors for many RV spectrographs, although the faint magnitudes of these stars implies that mass-constraining RV measurements will be very time-intensive.

The CARMENES \citep{Quirrenbach2010} instrument would require 460~s exposures to obtain 7.1 m~s$^{-1}$ precision for TOI 122 and 2250~s exposures to obtain 3.4 m~s$^{-1}$ precision for TOI 237b\footnote{\href{https://carmenes.caha.es/ext/instrument/index.html}{https://carmenes.caha.es/ext/instrument/index.html}}.
The latter is just beyond the 1800s maximum individual exposure time for this instrument, but the former implies the mass of TOI 122b could be within reach of a reasonably ambitious CARMENES observing program.
Likewise, the Habitable Zone Planet Finder (HPF) spectrograph \citep{Mahadevan2012,Mahadevan2014} could possibly achieve precision as good as 10 m s$^{-1}$ for TOI 122 and 5 m s$^{-1}$ for TOI 237 with 15-minute exposures \citep[see Fig. 2 of][]{Mahadevan2012}. With slightly longer exposure times, this instrument may be able to achieve mass-constraining precision for these two planets. The recent discovery of the G 9-40 system \citep{Stefansson2020AJ} used HPF to constrain planetary masses, achieving 6.49 m s$^{-1}$ precision with exposure times of 945~s. This star has K$_{\rm{s}}=9.2$, so scaled to the magnitudes of TOIs 122 and 237, we would need exposure times of $\sim$4~ks to achieve this precision for the systems presented here. Another instrument, the InfraRed Doppler (IRD) for the \textit{Subaru} telescope \citep{Kotani2014} also provides some hope. The sensitivity estimator\footnote{\href{http://ird.mtk.nao.ac.jp/IRDpub/sensitivity/sensitivity.html}{http://ird.mtk.nao.ac.jp/IRDpub/sensitivity/sensitivity.html}} implies that for both of these stars, $\sim$2 m s$^{-1}$ precision (S/N$>$100) may be possible with 1 hr exposures.

\subsection{Atmospheric Characterization Prospects}
In order to assess the viability of TOI 122b and TOI 237b for atmospheric studies, we calculated their emission spectroscopy metrics (ESM) following \citet{Kempton2018}. This metric represents the S/N of a single secondary eclipse observed by JWST's MIRI LRS instrument. The emission S/N scales directly as the flux of the planet and the square root of the number of detected photons, and inversely to the flux of the star, so hot planets orbiting cool nearby stars will have a larger ESM.

We calculate the ESM assuming that the planet dayside temperatures are equal to 1.1$\times$T$_{\rm{eq}}$ \citep[following the process outlined in][]{Kempton2018}, and that both have an Earth-like albedo of 0.3. We find ESM to be 2.9 for TOI 122b and 0.6 for TOI 237b. Compared to GJ 1132b (ESM $=$ 7.5) these planets are much less favorable for atmospheric follow-up with JWST. A minimum of 12 eclipses would be necessary to achieve a S/N $>$ 10 for TOI 122b and a minimum of 278 eclipses would be needed for TOI 237b, as the S/N scales as $\sqrt{N_{obs}}$. Detecting thermal emission with JWST would be challenging for TOI 122b and impractical for TOI 237b.

We also calculate the transmission spectroscopy metric (TSM) from \citet{Kempton2018}. This metric corresponds to the expected S/N of transmission features for a cloud-free atmosphere, over 10 hours of observation (5 hours in-transit). Our predicted TSMs are 54 for TOI 122b and 7 for TOI 237b, which imply these planets could both be amenable to transmission spectroscopy with JWST's NIRISS instrument, although planetary mass measurements would be necessary to make precise inferences from their transmission spectra \citep{Batahla2019}.

\begin{figure}[b!]
\centering
\includegraphics[width=0.47\textwidth]{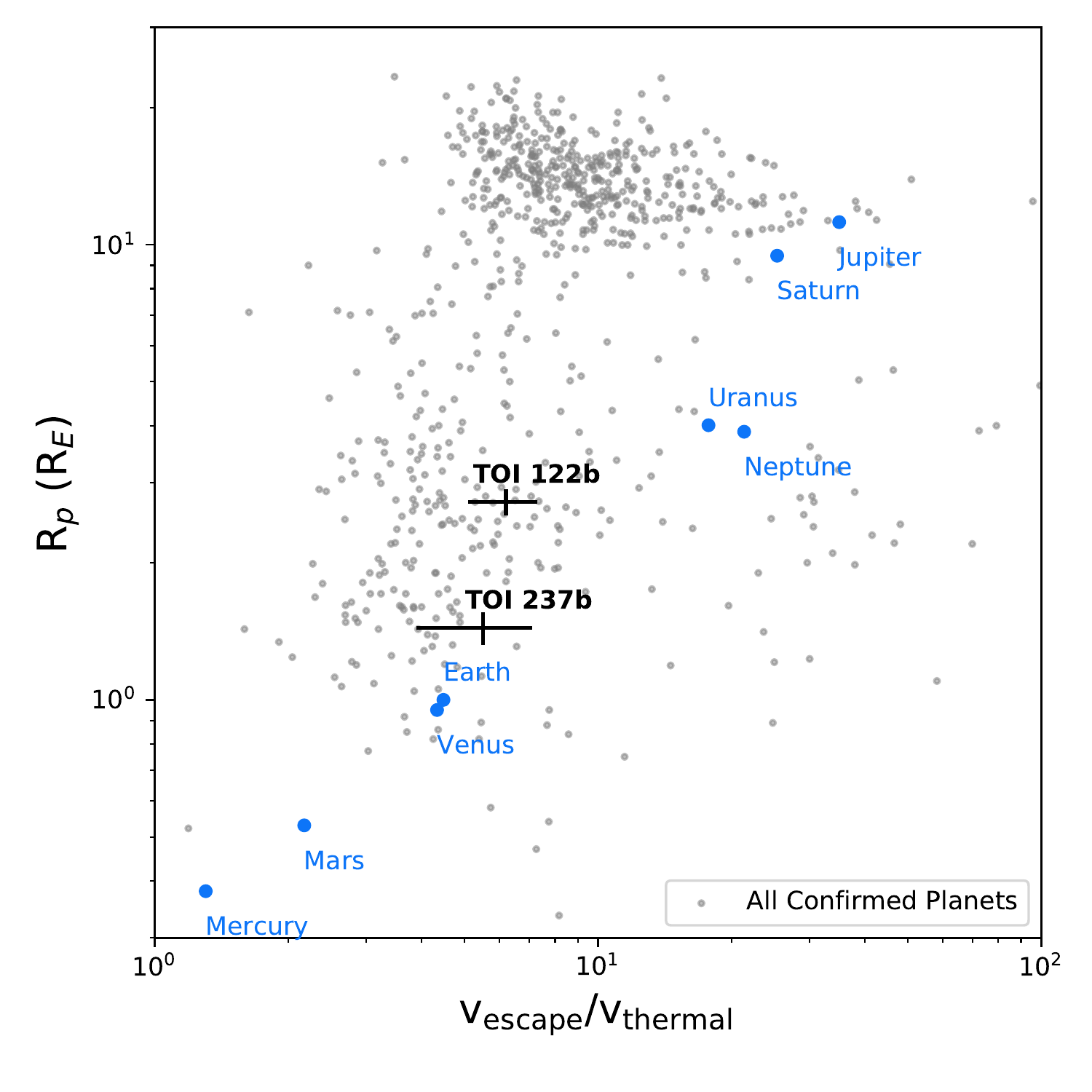}
\caption{The ratio of planetary escape velocity to the thermal energy of an H atom at the planetary equilibrium temperature \citep[the ``escape parameter'';][]{Jeans1905}, for known transiting and Solar System planets. TOI 122b and TOI 237b are included, using predicted masses from \citet{Chen2017} to calculate their gravity. This extremely rough proxy for susceptibility to atmospheric escape indicates these planets may be broadly similar to Earth and Venus, in terms of ongoing mass loss from their atmospheres. This qualitative comparison does not account for the important XUV radiation illuminating the planets, either now or in the past. We estimate the uncertainties for TOI 122b and 237b by propagating our uncertainties from the planet parameters, which are dominated by large uncertainties on predicted masses. \label{fig:vesc}}
\end{figure}

\subsection{Volatile Evolution}
These two planets span an interesting range of radii and insolations, making them exciting cases that may help us learn more about the diversity of atmospheres possessed by small planets orbiting M dwarfs. Figure \ref{fig:vesc} shows the Jeans escape parameter \citep[e.g.,][Box 2.2]{PlanetaryClimates} for these systems as well as Solar System bodies and all confirmed exoplanets for which this parameter could be calculated. This ratio of gravitational-to-thermal energy is an extremely approximate tracer of atmospheric escape, but it can help us qualitatively understand the relative susceptibility of different planets to atmospheric loss. With only loose predictions for the masses of TOI 122b and TOI 237b, their position on this plot leaves us with an ambiguous picture of whether they have atmospheres and what their compositions could be. They may even represent the transition between worlds that have lost almost all of their H/He (such as Earth and Venus) and worlds that have retained those lighter elements (such as Neptune or Uranus). Though we cannot determine any strong constraints with this Jeans approximation alone, these two planets are not in a regime where they would have obviously lost their atmospheres, as Mercury and Mars have. A more detailed investigation into the current and past XUV irradiation, which is a main driver of atmospheric loss, would be necessary to more cleanly place these planets in context \citep{Zahnle2017}. 

{\bf TOI 122b} is a sub-Neptune-sized planet orbiting an M dwarf that is 33\% the radius of our Sun. It likely has a thick atmosphere but on a 5.1 day orbit, it is far interior to the habitable zone of its star and irradiated at over 8$\times$ the flux of the Earth. It is dim enough to present a challenge for most existing radial velocity instruments, but mass measurements might be possible with a sufficient investment of time on IR spectrographs. Its atmosphere is on the edge of detectability in both emission and transmission with JWST. With a relatively low equilibrium temperature, there could be very interesting atmospheric chemistry in this planet's atmosphere that might be observable with sufficiently ambitious observing programs.

{\bf TOI 237b} is a super-Earth-sized planet orbiting a M dwarf that is 21\% the radius of our Sun and only 3200 K. With its 5.4 day orbit, it receives nearly 4$\times$ Earth insolation from its host star. Given the size of this planet and dimness of the star, mass measurements are likely very difficult to achieve, and we may not know its mass for some time. Even cooler than TOI 122b, this planet cannot be studied with emission spectroscopy, but transmission spectroscopy is possible and we may be able to learn about this planet's atmosphere, if it has retained one.

We are left with the following pictures of these systems: TOI 122b and TOI 237b are two worlds that span planetary radii not seen in our own solar system and are interesting laboratories to study planet formation, dynamics, and composition. Their long periods leave them too cool for emission spectroscopy but as a result, they occupy a very interesting space of relatively cool, though still uninhabitably warm, planets. Thus, they may give us insight to an as-yet poorly understood type of planetary atmosphere. While more targeted atmospheric or radial velocity studies would require a significant investment of time for these two systems, they are valuable additions to the statistical distribution of known planets.

\textbf{Software}
Python code used in this paper is available on the author's Github\footnote{\href{https://github.com/will-waalkes/TOI237and122}{https://github.com/will-waalkes/TOI237and122}}.This project made use of many publicly available tools and packages for which the authors are immensely grateful. In addition to the software cited throughout the paper, we also used \texttt{Astropy} \citep{astropy}, \texttt{NumPy} \citep{numpy}, \texttt{Matplotlib} \citep{matplotlib}, \texttt{Pandas} \citep{pandas}, and Anaconda's \texttt{JupyterLab}.

\textbf{Acknowledgements}
Funding for the TESS mission is provided by NASA's Science Mission directorate. We acknowledge the use of public TESS Alert data from pipelines at the TESS Science Office and at the TESS Science Processing Operations Center. This research has made use of the ExoFOP-TESS website, which is operated by the California Institute of Technology, under contract with the National Aeronautics and Space Administration under the Exoplanet Exploration Program. This paper includes data collected by the TESS mission, which are publicly available from the Mikulski Archive for Space Telescopes (MAST). This material is based upon work supported by the National Science Foundation Graduate Research Fellowship Program under Grant No. (DGE-1650115) and (DGE-1746045). Any opinions, findings, and conclusions or recommendations expressed in this material are those of the authors and do not necessarily reflect the views of the National Science Foundation. This work makes use of observations from the LCOGT network. Resources supporting this work were provided by the NASA High-End Computing (HEC) Program through the NASA Advanced Supercomputing (NAS) Division at Ames Research Center for the production of the SPOC data products. The research leading to these results has received funding from  the ARC grant for Concerted Research Actions, financed by the Wallonia-Brussels Federation. TRAPPIST is funded by the Belgian Fund for Scientific Research (Fond National de la Recherche Scientifique, FNRS) under the grant FRFC 2.5.594.09.F, with the participation of the Swiss National Science Fundation (SNF). MG and EJ are F.R.S.-FNRS Senior Research Associates. B.R-A. acknowledges the funding support from FONDECYT through grant 11181295.

\clearpage

\bibliography{main.bib}

\end{document}